\documentclass[aps,
twocolumn,reprint,
showpacs,preprintnumbers,
amsmath,amssymb,floatfix,a4paper,longbibliography]{revtex4-2} 
\usepackage{amsmath}
\usepackage{amsfonts}
\usepackage{amssymb}
\usepackage{bm}
\usepackage{color}
\usepackage{graphicx}

\usepackage{hyperref}
\usepackage{dcolumn}
\usepackage{braket}
\usepackage{ulem}
\usepackage{xspace}
\usepackage{siunitx}

\newcommand{\etal}{\textit{et al.}}

\newcommand{\omegal}{\omega_{l}}
\newcommand{\omegah}{\omega_{h}}
\newcommand{\omegam}{\omega_{m}}
\newcommand{\highmid}{hm}
\newcommand{\mh}{mh}
\newcommand{\lm}{lm}
\newcommand{\ml}{ml}
\newcommand{\hl}{hl}
\newcommand{\mm}{mm}
\newcommand{\FC}{\texttt{FC}}
\newcommand{\B}{\texttt{B}}
\newcommand{\PS}{\texttt{P}}
\newcommand{\ODB}{\texttt{ODB}}
\newcommand{\SWAP}{\texttt{SWAP}}
\newcommand{\nMode}{M}

\newcommand{\Ca}{\mathrm{{}^{40}Ca^+}}

\newcommand{\erf}{\mathrm{erf}}
\newcommand{\ii}{\mathrm{i}}
\newcommand{\ee}{\mathrm{e}}

\newcommand{\mum}{\unit{\micro\meter}}

\newcommand{\mus}{\unit{\micro\second}}

\newcommand{\kHz}{\mathrm{kHz}}
\newcommand{\MHz}{\mathrm{MHz}}
\newcounter{num}
\newcommand{\Rnum}[1]{\setcounter{num}{1} \Roman{num}}
\newcommand{\rnum}[1]{\setcounter{num}{1} \roman{num}}

\draft 

\begin{document}

\preprint{}

\title{A method of an on-demand beamsplitter for trapped-ion quantum computers} 

\author{Takanori Nishi}
\email{nishi@chem.s.u-tokyo.ac.jp}
\affiliation{Institute for Attosecond Laser Facility, The University of Tokyo, 7-3-1 Hongo, Bunkyo-ku, Tokyo 113-0033, Japan}
\date{\today}

\begin{abstract}
Quantum information processing using local modes of trapped ions has been applied to implementing bosonic quantum error correction codes and conducting efficient quantum simulation of bosonic systems.
However, control of entanglement among local modes remains difficult because entanglement among resonant local modes is governed by the Coulomb interaction, which is not switchable.
We propose a method of a beamsplitter for a trapped-ion architecture, where the secular frequency of each mode is dynamically controllable.
The neighboring modes are far detuned except when the beamsplitter needs to be applied to them.
We derive the analytical formula of the proposed procedure and numerically confirm its validity.

\end{abstract}


\maketitle 

\section{INTRODUCTION}\label{Sec:Intro}
Continuous variable (CV) encoding of quantum information processing (QIP) provides an efficient way of constructing quantum error correction such as the Gottesman--Kitaev--Preskill (GKP) codes \cite{GKP2001, Grimsmo2021, Nathan2025} and the cat codes \cite{CatCode1999, Putterman2025} as well as simulation of bosonic systems \cite{PorrasCirac2004BEC, Toyoda2013, KatzMonroe2023, MacDonell2021, Olaya-Agudelo2025}.
The initialization and the error-correction cycle of the GKP codes have been demonstrated in trapped ions \cite{Fluhmann2019, deNeeve2022} and in superconducting circuits \cite{Campagne2020, Sivak2023}.
For trapped ions, a universal gate set for the GKP code has been realized in Ref. \cite{Matsos2025}, where two GKP qubits are encoded using two local modes of a single ion.
All of the logical gates are implemented by modulating phases of sideband lasers.
Because a single ion can accommodate only three local modes, a novel scheme for entangling local modes of different ions needs to be developed.
However, the control of entanglement among local modes of different ions is difficult because the entanglement of local modes is generated by phonon hopping induced by the Coulomb interaction, i.e., the entangling gate cannot be turned off even if it is not needed.
Because the Hamiltonian of phonon hopping is equivalent to that of a beamsplitter \cite{Brown2011, Toyoda2015}, phonon hopping can be regarded as a ``non-on-demand beamsplitter.''

When we change the basis from local modes to normal modes by diagonalizing the Hamiltonian including quadratic terms of the Coulomb interaction \cite{James1998}, phonon hopping among normal modes is induced only via higher order terms of the Coulomb interaction \cite{Nie2009, Ding2017, Ding2017CrossKerr}.
If the normal mode frequency is tuned so that the higher order coupling is negligible, phonons can be coupled only optically and thus the on-demand beamsplitter has been demonstrated by using sideband lasers \cite{Chen2023}.

In order to achieve a large scale CV-TIQC using the normal modes, we need to interconnect modules of tens of ions.
However, photonic \cite{Inlek2017} or cavity-mediated \cite{Ramette2022} interconnection developed for discrete variable encoding is not applicable to CV encoding as the mode frequency is typically a few MHz or less.
For local modes, by taking advantage of highly accurate ion transport using quantum charge-coupled devices (QCCD) \cite{QuantinuumH2_2023}, interconnection of two modules separated by $10\,\mum$ has been demonstrated \cite{Akhtar2023}.

In order to control phonon hopping of local modes due to quadratic terms of the Coulomb interaction,
we can use the fact that the rate of hopping $\kappa$ depends on the distance between ions $d$ as $\kappa\propto d^{-3}$.
By storing the ions while keeping the distance between them sufficiently large, phonon hopping can be suppressed.
To induce phonon hopping, we can transport two ions in close proximity, deform the trap potential to accelerate phonon hopping, and separate them again.
Lau and James \cite{Lau&James2012} theoretically showed that the beamsplitter as well as Gaussian and non-Gaussian single-mode operations can be implemented by combining modulation of the trap potential and ion transport, and thus the universal gate set for CV encoding can be implemented without laser.
We note that the Lau--James method can be extended to normal modes by treating a chain as a single harmonic oscillator \cite{Harlander2011}; 
the normal modes of two ion chains can be entangled by bringing the chains closer and then separating them as demonstrated experimentally in \cite{Valentini2025}.

Phonon hopping into a given mode can be suppressed without changing the distance of ions if transition frequencies of the mode are far detuned from those of the other modes.
By illuminating an ion with laser light resonant with a red sideband, the internal and the motional states of the ion become coupled and the transition frequencies of the coupled state are shifted from those without laser light.
Therefore, hopping from the other modes into the illuminated mode is suppressed, which is called a phonon blockade \cite{Debnath2018,Ohira2021PhononBlockade}.
Because the state of the mode driven by the laser is disturbed unless it is in the ground state, the phonon blockade is only applicable to state preparation or state detection but not to gate operation.

We can also suppress phonon hopping without ion transport or coupling with internal states by using dynamical decouping (DD) \cite{Shen2014, Ohira2022DD}, where all the modes are resonant and phonon hopping proceeds among all the modes.
Control of hopping is achieved by applying a sequence of $\pi$-phase shift ($\PS_{\pi}$) gates to a set of modes so that undesirable hopping is canceled at the end of the sequence.
The fidelity of cancellation decreases as the duration of the $\PS_{\pi}$ gate increases because phonon hopping proceeds even during the $\PS_{\pi}$ gate, which degrades the fidelity of the $\PS_{\pi}$ gate.
A fast $\PS_{\pi}$ gate can be implemented by modulating the trap potential as demonstrated in Ref. \cite{Mielenz2016}, with which we are able to cancel phonon hopping and implement multi-mode entangling gates with high fidelity by a method called the cancellation of Coulomb coupling by modulating harmonic potential (C3PO) \cite{Nishi2025C3PO}.

In the present study, we propose an on-demand beamsplitter for local modes by controlling the detuning of each local mode using a frequency-changing method within a few microseconds.
The present method does not change the population of any modes unlike the phonon blockade or need any gate operation to cancel phonon hopping unlike the DD.
We derive the analytical formula of the procedure and confirm its validity by numerical simulation.
Because phonon hopping among different ions is suppressed except during the beamsplitter gate, the present method can reduce unwanted hopping during state preparation, single-mode gates, and state readout, which are crucial in encoding the GKP states and repeating error correction cycle.
We note that the implementation of a beamsplitter using a similar idea was demonstrated for the normal mode \cite{Ding2017, Ding2017CrossKerr}; the detuning between two normal modes are decreased by changing the DC voltages applied to the rods of the linear trap and are increased again after certain time to induce desired amount of hopping.
However, extension of the method to large number of normal modes is not clear because the change of the voltage induces frequency change of many normal modes in general.
On the other hand, the present method relies on the control technique \cite{Mielenz2016} which can modify the curvature of the electric potential at each ion's position independently.

\section{Method}\label{Sec:Theory}
\subsection{On-demand beamsplitter between local modes}
We consider a set of ions trapped along the axial direction $Z$ and assume that the secular frequency along the radial direction $X$ of each ion is a few MHz while the hopping rate is a few kHz or less.
Therefore, for a set of $\nMode$ ions, we can consider $\nMode$ local modes along the $X$ direction.
We assume that the ions are trapped by a surface electrode trap and the secular frequency  $\tilde{\omega}_j(t)$ of the $j$th mode can be controlled by modulating the DC voltage applied to the electrodes as demonstrated in \cite{Mielenz2016}.
Although the shape of the RF electrode in \cite{Mielenz2016} was optimized to trap three ions in a triangular geometry, the optimization method \cite{Schmied2009} adopted in \cite{Mielenz2016} is applicable to any trap geometry; by specifying the ion positions and the curvature tensors, the shape of the RF electrode is determined so that the curvature per voltage applied on the RF electrode is maximized.
We also note that the trap frequency of local modes can be tuned by applying optical tweezer to each ion \cite{Mazzanti2021}.
The maximum trap frequency achievable by tweezer was estimated as $2\pi\times70\,\kHz$ for $\Ca$ \cite{Mazzanti2021}.
In the present study, the amount of frequency change should be at least $2\pi\times200\,\kHz$ to implement the beamsplitter.
Therefore, we could adopt tweezer when the necessary frequency change is small, such as a phase shift gate, but the detailed analysis is left for future work.

The Hamiltonian of the set of $M$ ions with mass $m$ is given by
\begin{align}
    H(t)&= H_0 + \sum_j{\frac{\hbar\Omega_j^2(t)}{4\omega_j}(a^{\dagger}_{j,\omega_j} + a_{j,\omega_j})^2} \label{eq:Hamiltonian} \\
    H_0&=\sum_j\hbar\omega_j\left(a_{j,\omega_j}^{\dagger}a_{j,\omega_j} + \frac{1}{2}\right) \notag\\
    &\hspace{5mm}+\sum_{j>k}\frac{\hbar\kappa_{j,k}}{2}\left(a_{j,\omega_j}^{\dagger}a_{k,\omega_k}+a_{j,\omega_j}a_{k,\omega_k}^{\dagger}\right), \label{eq:H0}
\end{align}
where $H_0$ is the Hamiltonian of the set of ions without any modulation of the DC voltage, $\omega_j$ is the static secular frequency of the $j$th mode, $\kappa_{j,k}$ is the rate of hopping between the $j$th and the $k$th modes.
The frequency modulation is described by the quadratic term in equation \eqref{eq:Hamiltonian} and the difference of the squared frequency from the static secular frequency defined by $\Omega_j^2(t)\equiv \tilde{\omega}_j^2(t) - \omega_{j}^2$ satisfies $\Omega_j(0)=0$.
The lowering (raising) operator $a_{j,\omega_j}\, (a_{j,\omega_j}^{\dagger})$ of the $j$th mode is defined with respect to the $j$th coordinate scaled by $x_{0,j}=\sqrt{\hbar/(m\omega_j)}$.

First, we describe the scheme of the on-demand beamsplitter for $\nMode=2$ and then extend it to the general case. 
We assume that three values of secular frequency are available; lower and higher ``memory frequencies'', $\omegal$ and $\omegah$, respectively, and a ``gate frequency'' $\omegam$, where they satisfy $\omegal <\omegam < \omegah$ and the detunings, $\delta_{\highmid}=\omegah - \omegam$ and $\delta_{\ml}=\omegam - \omegal$, are sufficiently large so that phonon hopping is negligible between the modes oscillating at different frequencies.
The gate frequency $\omegam$ is used only for the implementation of the beamsplitter as we discuss later.
We choose the static secular frequencies as $\omega_0=\omegah$ and $\omega_1=\omegal$.

The entanglement between neighboring modes can be generated in three steps: (1) the frequency changing ($\FC$) of each mode to $\omegam$, which we denote as $\FC(\omega\xrightarrow{}\omegam)$, (2) phonon hopping, and (3) the inverse $\FC$ ($i\FC$) back to the memory frequency, $i\FC(\omegam\xrightarrow{}\omega)$.
For $M=2$, we first apply the downward $\FC$ to the 0th mode $\FC_0(\omegah\xrightarrow{}\omegam)$ and the upward $\FC$ to the 1st mode $\FC_1(\omegal\xrightarrow{}\omegam)$.
Then, because the resulting states are resonant, phonon hopping proceeds.
Finally, once the desired amount of entanglement is generated, we apply $i\FC_0(\omegam\xrightarrow{}\omegah)$ and $i\FC_1(\omegam\xrightarrow{}\omegal)$.
Unlike C3PO, the proposed scheme does not need to cancel unwanted entanglement and provides an entangling gate only when we need it, and thus we call the scheme the on-demand beamsplitter.

As we show in section \ref{subsec:simplification_FC}, the above procedure is equivalent to a beamsplitter sandwiched by additional phase-shift gates.
Because the phase-shift gate can be implemented by sequentially applying $\FC$ and $i\FC$ as shown in equation \eqref{eq:PS_equals_iFCandFC}, we can cancel the additional phase shift if needed by applying phase-shift gates before and after the on-demand beamsplitter.

\begin{figure}[htbp]
\begin{center}
\includegraphics[width=8.6cm]{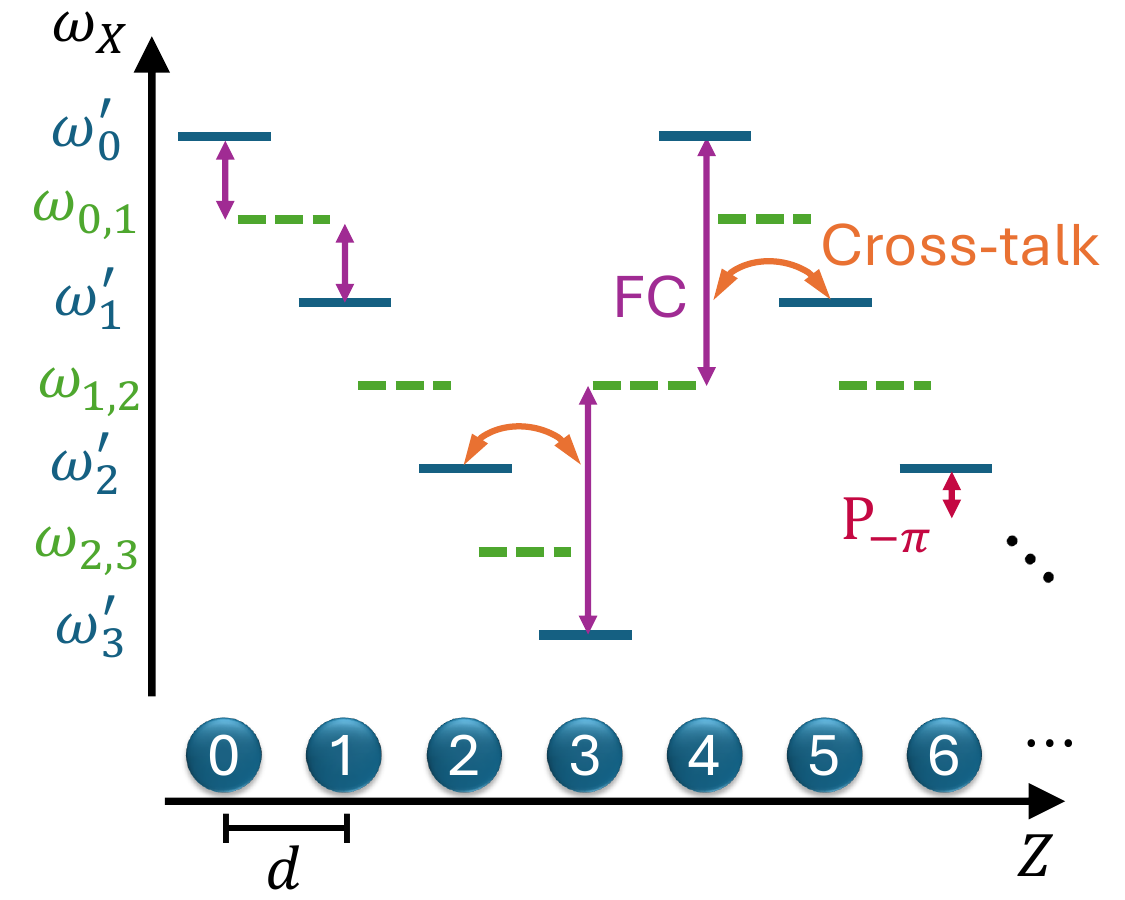}
\caption{Sawtooth configuration of frequencies for $N=4$.
Ions are aligned along $Z$ axis and separated by $d$.
Local modes along $X$ is used for computation.
A gap between a memory frequency $\omega'_{\alpha}\,(\alpha=0,1,\dots,N-1)$ (solid horizontal bar) and a gate frequency $\omega_{\beta-1,\beta}\,(\beta=1,2,\dots,N)$ should be large enough to avoid unwanted phonon hopping.
For the on-demand beamsplitter between $(j-1)$th and $j$th modes with $j\neq0\mod N$, the frequency of each mode is changed by $\FC$ (purple arrow) to the gate frequency $\omega_{\alpha,\beta}\,(\alpha=j-1,\,\beta=j\mod N)$ (dashed horizontal bar).
Cross-talk error occurs during the on-demand beamsplitter between $(j-1)$th and $j$th modes with $j=0\mod N$.
The largest contribution to the cross-talk error comes from $(j-2)$th mode, which becomes resonant with the $(j-1)$th mode, and from $(j+1)$th mode, which becomes resonant with the $j$th mode during $\FC$.
Phonon hopping between resonant modes separated by $Nd$ can be mitigated by C3PO, e.g., phonon hopping between the 2nd and the 6th modes can be mitigated by applying $(-\pi)$-phase shift gates $\PS_{-\pi}$ (red arrow) to the 6th mode.
$\PS_{-\pi}$ is implemented by $\FC$ and $i\FC$ but the amount of change in frequency can be smaller than that needed for the on-demand beamsplitter.
Note that $\PS_{-\pi}$ is equivalent to $\PS_{\pi}$ up to global phase and we choose either of them depending on the situation.}
\label{fig:sawtooth}
\end{center}
\end{figure}

For $\nMode>2$, we need to increase the number of the memory and the gate frequencies.
If $M$ memory frequencies and $M-1$ gate frequencies are available, each mode can be trapped at different frequency and the on-demand beamsplitter can be applied to the nearest-neighbor (NN) modes.
However, such an assumption becomes unrealistic for large $\nMode$ because the trap frequency generated by the surface electrode trap is generally about a few MHz while each frequency needs to be separated by a few hundreds of kHz to avoid phonon hopping.
Therefore, we consider limited number of frequencies while keep the distance between resonant modes as large as possible.

When $N$ memory frequencies $\omega'_0>\omega'_1>\dots>\omega'_{N-1}$ are available with $N<M$, we assign the $j$th mode frequency as $\omega_j=\omega'_{\alpha}$ with $j=\alpha\mod N$ for $j=0,1,\dots,\nMode-1$ and $\alpha=0,1,\dots,N-1$.
Gate frequencies are denoted as $\omega_{\alpha,\alpha+1}$ such that $\omega'_{\alpha}>\omega_{\alpha,\alpha+1}>\omega'_{\alpha+1}$ is satisfied.
As we have a large jump of frequency between $\omega_{j-1}$ and $\omega_j$ for $j=0\mod N$, we call this assignment of frequencies ``sawtooth configuration.''
An example for $N=4$ is shown in figure \ref{fig:sawtooth}.
When the distance between the NN modes is given by $d_{j,j+1}=d,\,\forall j$, the distance between resonant modes is $Nd$.
Because the hopping rate decreases with the distance as $\kappa_{j,k}\propto d_{j,k}^{-3}(\sqrt{\omega_j\omega_k})^{-1}$, the hopping between resonant modes during the operation of the on-demand beamsplitter between the NN modes becomes negligible as $N$ increases.

In order to avoid inducing hopping between two modes whose distance is smaller than $Nd$, the gate frequency $\omega_{\alpha,\alpha+1}$ needs to be well separated from any memory frequency.
For example, for the $j$th mode whose frequency is given by $\omega_j=\omega'_{\alpha+1}$, we consider the upward $\FC$ to $\omega_{\alpha,\alpha+1}$.
Then, if $\omega_{\alpha,\alpha+1}$ is not separated well from $\omega'_{\alpha}$, hopping proceeds between $j$th and $(j+N-1)$th modes because $\omega_{j+N-1}=\omega'_{\alpha}$.

The scheme explained for $M=2$ can be applied to the general $M$ with a sawtooth configuration, however, two types of errors should be taken into account.
(1) The ``cross-talk error'' during $\FC$ of the on-demand beamsplitter between $(j-1)$th and $j$th modes with $j=0\mod N$.
Because $\omega_{j-1}=\omega'_{N-1}\equiv\omegal$ and $\omega_{j}=\omega'_{0}\equiv\omegah$, there exists at least one memory frequency $\omega_{k}$ satisfying $\omegal<\omega_k<\omegam$ or $\omegam<\omega_k<\omegah$ for any gate frequency $\omegam$.
Therefore, the modes oscillating at $\omega_k$ become resonant for short time with the $(j-1)$th mode during $\FC(\omegal\xrightarrow{}\omegam)$ or with the $j$th mode during $\FC(\omegah\xrightarrow{}\omegam)$.
This type of error becomes small as the duration of $\FC$ becomes short.
(2) Hopping between resonant modes separated by $Nd$. 
This type of error can be mitigated by applying C3PO. Because the hopping rate is much smaller than that for the NN modes, the number of $\PS_{\pi}$ gates needed to cancel the hopping can be much smaller than that shown in \cite{Nishi2025C3PO}, where the NN modes are also involved.
Note that $\PS_{\pi}$ is equivalent to $\PS_{-\pi}$ up to an irrelevant global phase and we use either of them depending on the situation.
We can also consider a ``shifted sawtooth configuration'', where the static frequency of the $j$th and the $(j+kN)$th modes are tuned as $\omega_{j+kN}=\omega_j$ for even $k$ and $\omega_{j+kN}=\omega_j+\delta_j$ for odd $k$. 
The shift $\delta_j$ can be chosen as, e.g., $\delta_1=(\omega_{0,1} - \omega'_1)/2$ in figure \ref{fig:sawtooth} so that $\omega_5$ is detuned from $\omega_1$ as well as from the gate frequency $\omega_{0,1}$.
The requirement to implement the original sawtooth structure in figure \ref{fig:sawtooth} is the ability of the frequency change in the range $\omega_j\in[\omega'_3, \,\omega'_0]$.
Therefore, for $\delta_j>0\,\forall j$, the shifted sawtooth configuration can be implemented once the frequency change in the range $\omega_{kN}\in[\omega_0,\omega_0+\delta_0]$ for odd $k$ is available as well as the original requirement is fulfilled.
The effect of these two errors are investigated numerically in section \ref{subsec:SWAP_GKP_4modes}.

\subsection{Time evolution in the Heisenberg picture}
First, we describe the time evolution during $\FC$ in the Heisenberg picture by using the Lewis--Riesenfeld theory \cite{LewisRiesenfeld1969} (see Appendix \ref{app:LR_theory}).
We consider a time-dependent harmonic oscillator (TDHO) defined by the Hamiltonian $H(t)=P^2/(2m) + m\tilde{\omega}^2(t) X^2/2$ and $\FC$ from $\tilde{\omega}(t_i)=\omega_i$ to $\tilde{\omega}(t_f)=\omega_f$.
The time evolution of the TDHO can be described using an auxiliary function $b(t)$, which is related to $\tilde{\omega}(t)$ as
\begin{align}
    \ddot{b}+\tilde{\omega}^2(t)b-\frac{\omega_i^2}{b^3}&=0. \label{eq:omega(t)_and_b(t)}
\end{align}
The lowering and the raising operators of the initial state are denoted by $\bm{a}_{\omega_i}=[a_{\omega_i},\,a^{\dagger}_{\omega_i}]^{\top}$.
The operators at $t_f$ in a coordinate scaled by $\sqrt{\hbar/(m\omega_f)}$, $\bm{a}_{\omega_f}(t_f)=[a_{\omega_f}(t_f),\,a^{\dagger}_{\omega_f}(t_f)]^{\top}$, can be related to $\bm{a}_{\omega_i}$ using the transformation matrix $S_{if}$ as $\bm{a}_{\omega_f}(t_f)=S_{if}\bm{a}_{\omega_i}$, where 
\begin{align}
    \label{eq:Smatrix_af_by_ai}
    S_{if} =
    \begin{bmatrix}
        \eta^*e^{\ii\Theta(t_f)} & -\zeta e^{-\ii\Theta(t_f)}\\   
        -\zeta^* e^{\ii\Theta(t_f)} & \eta e^{-\ii\Theta(t_f)}
    \end{bmatrix},
\end{align}
where $\Theta$, $\eta$, and $\zeta,$ are defined using $(\omega_i,\,\omega_f,\,b(t))$ through \eqref{eqApp:Theta}, \eqref{eqApp:eta}, and \eqref{eqApp:zeta}, respectively.

In the following, we consider two ions with $\omega_0=\omegah$ and $\omega_1=\omegal$ and describe the time evolution of  $\bm{a}_{hl}\equiv[a_{0,\omegah},\,a^{\dagger}_{0,\omegah}, a_{1,\omegal},\,a^{\dagger}_{1,\omegal}]^{\top}$ during the on-demand beamsplitter in three steps; (1) $\FC$ to $\omegam$ from $t=0$ to $t=T_1=T_{\FC}$, (2) phonon hopping for the duration $T_{\B}$, i.e., from $t=T_1$ to $t=T_2=T_{\FC}+T_{\B}$, and (3) $i\FC$ from $t=T_2$ to $t=T_3=2T_{\FC}+T_{\B}$.

In the first step, we denote the initial and the final frequencies by the subscript of $\tilde{\omega}(t)$ as well as the corresponding $b(t)$, e.g., $\tilde{\omega}_{\highmid}(t)$ and $b_{\highmid}(t)$ for $\FC_0(\omegah\xrightarrow{}\omegam)$.
By neglecting phonon hopping, the operators at $T_1$ in a coordinate scaled by $\sqrt{\hbar/(m\omegam)}$, $\bm{a}_{mm}(T_1)\equiv[a_{0,\omegam}(T_1),\,a^{\dagger}_{0,\omegam}(T_1), a_{1,\omegam}(T_1),\,a^{\dagger}_{1,\omegam}(T_1)]^{\top}$, are obtained as $\bm{a}_{\mm}(T_1)=S_{\FC, \highmid,\lm}\bm{a}_{\hl}$, where the transformation matrix becomes block diagonal as
\begin{align}
    S_{\FC, \highmid,\lm}&=
    \begin{bmatrix}
        S_{\highmid} & O\\   
        O & S_{\lm}
    \end{bmatrix}.\label{eq:Smatrix_for_FC}
\end{align}
The matrices $S_{\highmid}$ and $S_{\lm}$ are obtained by substituting $(\omega_i, \,\omega_f,\,b)=(\omegah,\,\omegam,\,b_{\highmid})$ or $(\omegal,\,\omegam,\,b_{\lm})$, respectively, in \eqref{eq:Smatrix_af_by_ai}.

The effect of hopping during $\FC$ can be relevant when the time-dependent frequencies of the two modes become close.
However, such an effect can be suppressed by setting the duration $T_{\FC}$ much smaller than the hopping time scale $\propto\kappa_{0,1}^{-1}$
and therefore we expect that equation \eqref{eq:Smatrix_for_FC} can well approximate the time evolution under equation \eqref{eq:Hamiltonian}.

In the second step, we keep the frequency of both modes at $\tilde{\omega}_0(t)=\tilde{\omega}_1(t)=\omegam$ for the duration $T_{\B}$ to induce the desired amount of hopping, during which the Hamiltonian becomes simple as
\begin{align}
    \label{eq:Hamiltonian_omega_mid}
    H(t)&\equiv H_{\B} \notag\\
    &=\sum_{j=0,1}\hbar\omegam\left(a_{j,\omegam}^{\dagger}a_{j,\omegam} + \frac{1}{2}\right) \notag\\
    &\hspace{5mm}+\frac{\hbar\tilde{\kappa}_{0,1}}{2}\left(a_{0,\omegam}^{\dagger}a_{1,\omegam}+a_{0,\omegam}a_{1,\omegam}^{\dagger}\right),
\end{align}
where $\tilde{\kappa}_{0,1}=e^2/(4\pi\epsilon_0d_{0,1}^3m\omegam)$ is the hopping rate at the gate frequency $\omegam$.
The time evolution with this Hamiltonian is well known.
In the interaction frame (IF) with respect to $\sum_{j}\hbar\omegam\left(a_{j,\omegam}^{\dagger}a_{j,\omegam} + 1/2\right)$,
the propagator $\exp{(-\ii H_{\B}^{\mathrm{IF}}T_{\B}/\hbar)}$ coincides with the unitary operator representing a beamsplitter $U_{\B}=\exp{[-\ii\theta(a_0^{\dagger}a_1+a_0a_1^{\dagger})]}$ with $\theta=\tilde{\kappa}_{0,1}T_{\B}/2$.
Therefore, the output of the second step in the IF is given by $\bm{a}^{\mathrm{IF}}_{\mm}(T_2)=B(\theta)\bm{a}_{\mm}(T_1)$, where
\begin{align}
    \label{eq:a_after_B}
    B(\theta)=
    \begin{bmatrix}
        \cos\theta & 0 & -\ii\sin\theta & 0 \\   
        0 & \cos\theta & 0 & \ii\sin\theta \\
        -\ii\sin\theta & 0 & \cos\theta & 0 \\
        0 & \ii\sin\theta & 0 & \cos\theta
    \end{bmatrix}.
\end{align}
In order to take into account the dynamical phase, we define the transformation matrix for the phase-shift gates on two modes as
\begin{align}
    \label{eq:Smatrix_phase_shift_definition}
    P(\phi_1,\,\phi_2)=
    \begin{bmatrix}
        e^{-\ii\phi_1} & 0 & 0 & 0 \\   
        0 & e^{\ii\phi_1} & 0 & 0 \\
        0 & 0 & e^{-\ii\phi_2} & 0 \\
        0 & 0 & 0 & e^{\ii\phi_2}
    \end{bmatrix},
\end{align}
and obtain $\bm{a}_{\mm}(T_2)=P(\omegam T_{\B},\omegam T_{\B})\bm{a}^{\mathrm{IF}}_{\mm}(T_2)$.

In the third step, we apply $i\FC$ as $\bm{a}_{\hl}(T_3)=S_{\FC, \mh,\ml}\bm{a}_{\mm}(T_2)$, where
\begin{align}
    \label{eq:Smatrix_for_iFC}
    S_{\FC, \mh,\ml} =
    \begin{bmatrix}
        S_{\mh} & O\\   
        O & S_{\ml}
    \end{bmatrix}.
\end{align}
The matrices $S_{\mh}$ and $S_{\ml}$ are defined similarly as $S_{\highmid}$ and $S_{\lm}$ of \eqref{eq:Smatrix_for_FC}.

\subsection{Phase shift and squeezing from the frequency changing} \label{subsec:simplification_FC}
The transformation matrix \eqref{eq:Smatrix_af_by_ai} is derived by imposing the initial conditions $b(t_i)=1$ and $\dot{b}(t_i)=0$, which corresponds to the requirements $\tilde{\omega}(t_i)=\omega_i$ and $\tilde{\omega}(t)$ being smooth at $t_i$.
In addition, we impose $b(t_f)=\sqrt{\omega_i/\omega_f}$ and $\dot{b}(t_f)=0$, which corresponds to $\tilde{\omega}(t_f)=\omega_f$ and $\tilde{\omega}(t)$ being smooth at $t_f$.
These conditions lead to $\eta(t_f)=1$ and $\zeta(t_f)=0$, with which the transformation matrix \eqref{eq:Smatrix_af_by_ai} is simplified as
\begin{align}
    \label{eq:Smatrix_eta1_zeta0}
    S_{if} =
    \begin{bmatrix}
        \ee^{\ii\Theta(t_f)} & 0\\   
        0 & \ee^{-\ii\Theta(t_f)}
    \end{bmatrix}\equiv P_{if}(\Theta(t_f)),
\end{align}
where we introduced $P_{if}$ to emphasize that the transformation matrix is the combination of a basis change and a phase shift:
in the Schr\"odinger picture, equation \eqref{eq:Smatrix_eta1_zeta0} implies that the population at $t_i$ in the initial Fock basis is equal to that at $t_f$ in the final Fock basis but the phase is acquired as $\ket{n;\omega_i}\mapsto\ee^{\ii(n+1/2)\Theta(t_f)}\ket{n;\omega_f}$.
Note that $P$ without a subscript can also be denoted as $P_{ii}$ since it does not change the basis.

On the other hand, if we describe the output of $\FC$ in terms of the coordinate scaled by $\sqrt{\hbar/(m\omega_i)}$ using equation \eqref{eqApp:a_basis_change}, equation \eqref{eq:Smatrix_eta1_zeta0} leads to
\begin{align}
    \label{eq:ai(tf)}
    a_{\omega_i}(t_f)
    &=a_{\omega_i} e^{\ii\Theta(t_f)}\cosh{r}- a^{\dagger}_{\omega_i}e^{-\ii\Theta(t_f)} \sinh{r},
\end{align}
where $r=\log{\sqrt{\omega_f/\omega_i}}$.
This means that, when we describe both the initial and the final states using the initial Fock basis $\ket{n;\omega_i}$, $\FC$ is the combination of a phase shift and squeezing with $r$ the squeezing strength.

The unitary operator corresponding to equation \eqref{eq:Smatrix_eta1_zeta0} can be written as
\begin{align}
    \label{eq:FC_operator_as_pureFC_and_phase_shift}
    U_{\FC}(\Theta(t_f)) =\sum_n  e^{\ii\Theta(t_f)(n+1/2)} \ket{n;\omega_f}\bra{n;\omega_i}.
\end{align}
It is well known that such a population transfer can be realized if the modulation of the Hamiltonian is slow so that the adiabatic approximation is valid.
However, because we have not made any assumption on the time scale of $\FC$, the duration $T_{\FC}$ can be arbitrarily small as long as $\tilde{\omega}(t)$ defined by equation \eqref{eqApp:omega(t)_by_b(t)} is real and $b(t)$ and $\dot{b}(t)$ satisfy the above boundary conditions.

We define the extension of equation \eqref{eq:Smatrix_eta1_zeta0} to the two modes by $P_{\alpha\beta,\gamma\delta}(\Theta_1,\Theta_2)\equiv P_{\alpha\beta}(\Theta_1)\oplus P_{\gamma\delta}(\Theta_2)$. 
Then, the output of the on-demand beamsplitter can be obtained as
\begin{align}
    \label{eq:Smatrix_all_Hesenberg_pic}
    \bm{a}_{\hl}(T_3)&=S_{\FC, \mh,\ml}P(\omegam T_{\B},\, \omegam T_{\B})B(\theta)S_{\FC, \highmid,\lm} \bm{a}_{\hl} \notag\\
    &=P_{\mh,\ml}(-\Theta_{\mh},\,-\Theta_{\ml})P_{mm,mm}(\omegam T_{\B},\, \omegam T_{\B})\notag\\
    &\hspace{5mm}\times B(\theta)P_{\highmid,\lm}(-\Theta_{\highmid},\,-\Theta_{\lm})\bm{a}_{\hl} \notag\\
    &=P_{\mh,\ml}(-\Theta_{\mh}+\omegam T_{\B},\,-\Theta_{\ml}+\omegam T_{\B})B(\theta) \notag\\
    &\hspace{5mm}\times P_{\highmid,\lm}(-\Theta_{\highmid},\,-\Theta_{\lm})\bm{a}_{\hl},
\end{align}
where we used $P_{if}(\Theta_1)P_{ii}(\Theta_2)=P_{if}(\Theta_1+\Theta_2)$.
In the IF with respect to $\sum_j{\hbar\omega_j(a_{j,\omega_j}^{\dagger}a_{j,\omega_j}+1/2)}$, we obtain
\begin{align}
    \label{eq:final_result_interaction_frame}
    \bm{a}^{\mathrm{IF}}_{\hl}(T_3) = P_{hh,ll}(-\omegah T_3, -\omegal T_3) \bm{a}_{\hl}(T_3).
\end{align}
The transformation matrix for the entire process in the IF is given as
\begin{align}
    \label{eq:Smatrix_all_interaction_frame}
    &S_{\ODB}^{\mathrm{IF}}(\theta) \notag\\
    &= P_{\mh,\ml}(-\Theta_{\mh}+\omegam T_{\B}-\omegah T_3, -\Theta_{\ml}+\omegam T_{\B}-\omegal T_3) \notag\\
    &\hspace{5mm}\times B(\theta)P_{\highmid,\lm}(-\Theta_{\highmid},\,-\Theta_{\lm}),
\end{align}
where we used $P_{ff}(\Theta_1)P_{if}(\Theta_2)=P_{if}(\Theta_1+\Theta_2)$.
Finally, the phase can be compensated by applying the phase shift at the beginning and the end as
\begin{align}
    \label{eq:Smatrix_phase_compensated}
    &P_{hh,ll}(\Theta_{\mh}-\omegam T_{\B}+\omegah T_3, \Theta_{\ml}-\omegam T_{\B}+\omegal T_3)\notag\\
    &\hspace{5mm}\times S_{\ODB}^{\mathrm{IF}}(\theta)P_{hh,ll}(\Theta_{\highmid},\,\Theta_{\lm}) \notag\\
    &= P_{\mh,\ml}(0,0)B(\theta)P_{\highmid,\lm}(0,0),
\end{align}
where $P_{if}(0,0)$ $(i,f\in\{h,m,l\})$ is the frequency change $\omega_i\mapsto\omega_f$ without phase shift.

\section{Results} \label{sec:Results}
We consider the sawtooth configuration with $N=4$ in figure \ref{fig:sawtooth}.
First, we demonstrate the on-demand beamsplitter between the 0th and the 1st modes by numerical simulation for two cases; the 50:50 beamsplitter in section \ref{subsec:HOM} and the SWAP gate in section \ref{subsec:SWAP_GKP}.
Then, we also simulate the SWAP gate at the large jump, i.e., between the 3rd and the 4th modes in section \ref{subsec:SWAP_GKP_4modes}.
We solve the time-dependent Schr\"odinger equation (TDSE) numerically using time-dependent variational principle with matrix-product state (TDVP-MPS) \cite{Haegeman2011,Haegeman2016}, 
which we implement using a Python package \texttt{mpsqd} \cite{mpsqd}.
Although we use the second quantization in deriving the analytical expressions,
we adopt the first quantization for the numerical simulation because it is straightforward to describe the GKP states.
The state vectors are represented in the scaled coordinate $x$ for each mode
and the operators are represented using $x$ and $p$ such as equations \eqref{eqApp:RWA_Hamiltonian_scaled_xp} and \eqref{eqApp:Hamiltonian_TDHO_1st}.
In sections \ref{subsec:HOM} and \ref{subsec:SWAP_GKP}, 
the number of grid points and the maximum of the bond dimension are 128 and 50, respectively, while in section \ref{subsec:SWAP_GKP_4modes} they are 64 and 20 to reduce computational cost.

\begin{figure}[htbp]
\begin{center}
\includegraphics[width=8.6cm]{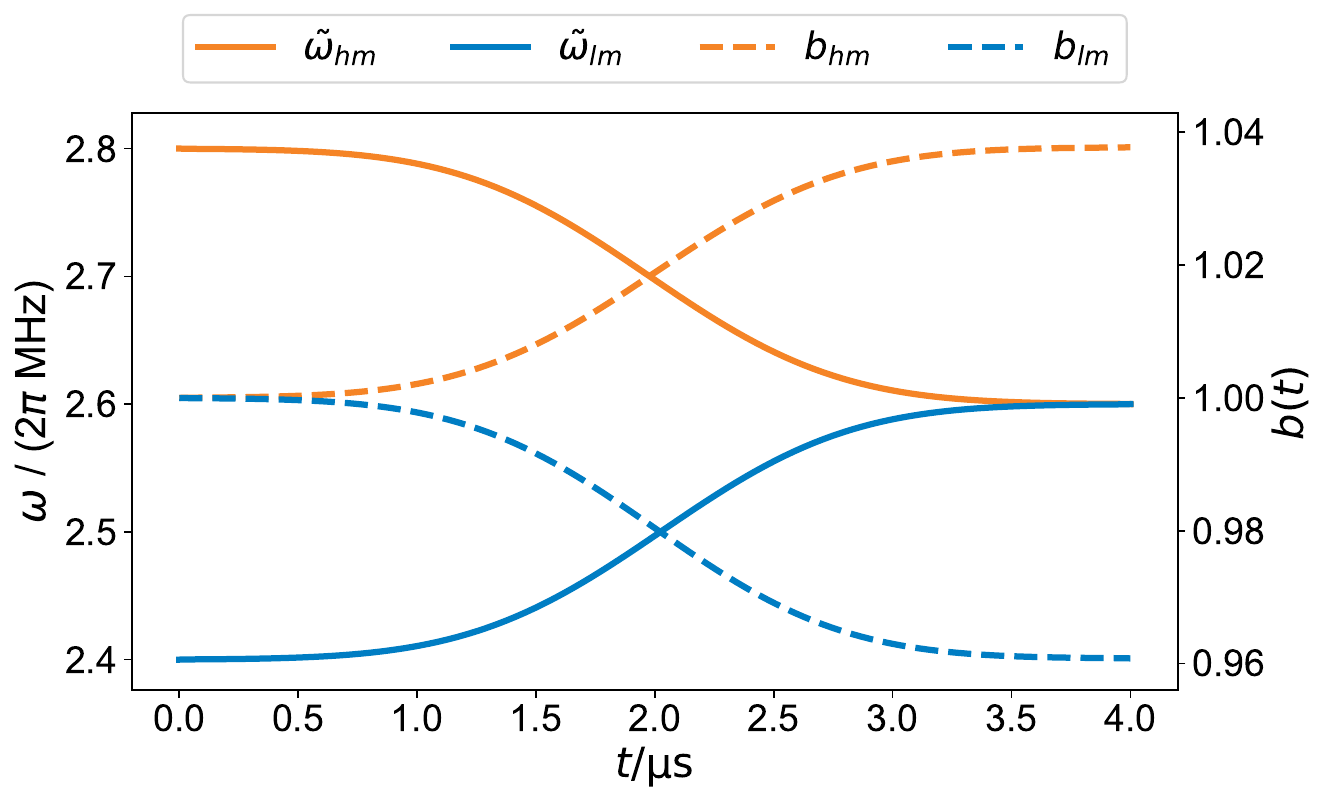}
\caption{The auxiliary function $b(t)$ (dashed lines) for the upward (blue) and the downward (orange) frequency changing defined by equations \eqref{eq:fu_erf} and \eqref{eq:fd_erf}, respectively.
The corresponding $\tilde{\omega}(t)$ (solid lines) is defined by equation \eqref{eqApp:omega(t)_by_b(t)}.}
\label{fig:b_and_omega}
\end{center}
\end{figure}

We consider $\Ca$ ions separated by $d=43.8\,\mum$ along $Z$ axis and we use their local modes along $X$ direction.
The memory frequencies are given by 
$\omega'_{\alpha}/(2\pi)=2.8-0.4\alpha\,\MHz\,(\alpha=0,1,2,3)$ and the gate frequencies by $\omega_{\beta,\beta+1}/(2\pi)=2.6-0.4\beta\,\MHz\,(\beta=0,1,2)$.
In sections \ref{subsec:HOM} and \ref{subsec:SWAP_GKP}, we only consider two ions, whose mode frequencies are 
$\omega_0=\omega'_0=2\pi\times2.8\,\MHz$ and $\omega_1=\omega'_1=2\pi\times2.4\,\MHz$, respectively.
The gate frequency is $\omega_{0,1}=2\pi\times2.6\,\MHz$, which corresponds to the hopping rate $\tilde{\kappa}_{0,1}=2\pi\times0.402\,\kHz$.
By simulating the static Hamiltonian defined by equation \eqref{eq:H0}, we have confirmed that hopping between two modes is negligible if $\omega_j\neq\omega_k$ and $\omega_j,\,\omega_k\in\{\omega'_0,\,\omega'_1,\,\omega_{0,1}\}$, and then we set $\omegah=\omega'_0$, $\omegal=\omega'_1$, and $\omegam=\omega_{0,1}$ for $\FC$.
The auxiliary function $b(t)$ is defined as
$b(t)=f_u(1-\sqrt{\omega_i/\omega_f},t)$ for the upward $\FC$ ($\omega_i < \omega_f$) while $b(t)=f_d(\sqrt{\omega_i/\omega_f}-1,t)$ for the downward $\FC$ ($\omega_i > \omega_f$), where
\begin{align}
    f_u(g,t)&=1-\frac{g}{2}\left(1+\erf\left[\left(\frac{t}{T_{\FC}}-\frac{1}{2}\right)\sigma\right]\right) \label{eq:fu_erf} \\
    f_d(g,t)&=1+\frac{g}{2}\left(1+\erf\left[\left(\frac{t}{T_{\FC}}-\frac{1}{2}\right)\sigma\right]\right) \label{eq:fd_erf}.
\end{align}
With this definition of $b(t)$, the phases for the upward $\FC$ and the downward $\FC$ for each mode are the same, i.e., $\Theta_{\highmid}=\Theta_{\mh}$ and $\Theta_{\lm}=\Theta_{\ml}$.
In sections \ref{subsec:HOM} ad \ref{subsec:SWAP_GKP}, we set the duration of $\FC$ to $T_{\FC}=4\,\mus$ and the width of the error function to $\sigma=6$.
Thus, the rate of $\FC$ in the present study is $(\omegah-\omegam)/T_{\FC}=2\pi\times 50\,\kHz/\mus$,
which is comparable to the rate $2\pi\times57\,\kHz/\mus$ demonstrated in \cite{Mielenz2016}.
Figure \ref{fig:b_and_omega} shows $\tilde{\omega}(t)$ and $b(t)$ for $\FC_0(\omegah\xrightarrow{}\omegam)$ and $\FC_1(\omegal\xrightarrow{}\omegam)$.

We note that these parameters satisfy the adiabatic condition as shown in appendix \ref{app:adiabatic_condition}.
However, because we derived the analytical formula \eqref{eq:Smatrix_all_interaction_frame} without adiabatic approximation, the on-demand beamsplitter can be implemented with much faster modulation of $\omega(t)$, e.g., with shorter $T_{\FC}$ and/or larger $g$.

In order to analyze the time evolution of phonon hopping in the Schr\"odinger picture, we use the following relation,
\begin{align}
    \label{eq:B_Schrodinger_pic}
    U_{\B}(\theta)\ket{j}_1\ket{k}_0    &=U_{\B}\left(a_1^{\dagger}\right)^j\left(a_0^{\dagger}\right)^k U_{\B}^{\dagger}U_{\B}\ket{0}_1 \ket{0}_0 \notag\\
    &=\left(a_1^{\dagger}\cos{\theta} -\ii a_0^{\dagger}\sin{\theta}\right)^j \notag\\
    &\times\left(-\ii a_1^{\dagger}\sin{\theta} + a_0^{\dagger}\cos{\theta}\right)^k \ket{0}_1 \ket{0}_0.
\end{align}
From equation \eqref{eq:FC_operator_as_pureFC_and_phase_shift}, we define the unitary operators for the two-mode $\FC$ and $i\FC$ as
\begin{align}
    U_{\FC}(\phi_0,\phi_1) &=U_{\FC_1}(\phi_1)U_{\FC_0}(\phi_0)\notag\\
    &=\sum_j  e^{\ii\phi_1(j+1/2)} \ket{j;\omegam}_1\bra{j;\omegal}_1\notag\\
    &\hspace{5mm}\otimes \sum_k  e^{\ii\phi_0(k+1/2)} \ket{k;\omegam}_0\bra{k;\omegah}_0 \label{eq:unitary_FC_2modes}\\
    U_{i\FC}(\phi_0,\phi_1)
    &=\sum_j  e^{\ii\phi_1(j+1/2)} \ket{j;\omegal}_1\bra{j;\omegam}_1\notag\\
    &\hspace{5mm}\otimes \sum_k  e^{\ii\phi_0(k+1/2)} \ket{k;\omegah}_0\bra{k;\omegam}_0,\label{eq:unitary_iFC_2modes}
\end{align}
respectively.
From equations \eqref{eq:Smatrix_all_interaction_frame}, \eqref{eq:B_Schrodinger_pic}, \eqref{eq:unitary_FC_2modes}, and \eqref{eq:unitary_iFC_2modes}, the unitary operator corresponding to the on-demand beamsplitter under the conditions $\Theta_{\highmid}=\Theta_{\mh}$ and $\Theta_{\lm}=\Theta_{\ml}$ is given by
\begin{align}
    \label{eq:U_ODB}
   &U_{\ODB}(\theta) \notag\\
   &=U_{i\FC}(-\Theta_{\highmid}+\omegam T_{\B}-\omegah T_3, 
    -\Theta_{\lm}+\omegam T_{\B}-\omegal T_3)\notag\\
    &\hspace{3mm}\times U_{\B}(\theta)U_{\FC}(-\Theta_{\highmid},-\Theta_{\lm}).
\end{align}
The two-mode phase-shift gate is defined as
\begin{align}
    \label{eq:unitary_phase_shift_2modes}
    U_{\PS}(\phi_0, \phi_1)&=U_{\PS_0}(\phi_0)\otimes U_{\PS_1}(\phi_1)\notag \\
    &=\sum_j  e^{-\ii\phi_1(j+1/2)} \ket{j;\omega}_1\bra{j;\omega}_1\notag\\
    &\hspace{5mm}\otimes \sum_k  e^{-\ii\phi_0(k+1/2)} \ket{k;\omega'}_0\bra{k;\omega'}_0,
\end{align}
where $\omega$ and $\omega'$ are the frequencies of the states $U_{\PS}$ acts on.
We note that the kets are labeled from the right while the parameters of the two-mode gates in equations \eqref{eq:Smatrix_phase_shift_definition}, \eqref{eq:unitary_FC_2modes}, \eqref{eq:unitary_iFC_2modes}, and \eqref{eq:unitary_phase_shift_2modes} are labeled from the left.
We also note that the phase-shift gate can be constructed by sequentially applying $\FC$ and $i\FC$, e.g., 
\begin{align}
    \label{eq:PS_equals_iFCandFC}
    U_{\PS_0}(\phi_a+\phi_b) &= \sum_k  e^{-\ii(\phi_a+\phi_b)(k+1/2)} \ket{k;\omegah}_0\bra{k;\omegah}_0 \notag\\
    &=U_{i\FC_0}(-\phi_b)U_{\FC_0}(-\phi_a).
\end{align}
Because the phase in equation \eqref{eq:PS_equals_iFCandFC} is acquired through cyclic and non-adiabatic evolution, it contains the geometric phase known as the Aharonov--Anandan (AA) phase \cite{AharonovAnandan1987, Zhang2023}.
The AA phase is obtained by subtracting the dynamical phase $-\hbar^{-1}\int_0^T dt\braket{\psi_k(t)|H(t)|\psi_k(t)}$ from the total phase $-(\phi_a+\phi_b)(k+1/2)$, where $T$ is the duration of $U_{\PS_0}(\phi_a+\phi_b)$ and $\ket{\psi_k(t)}$ is obtained by evolving $\ket{\psi_k(0)}=\ket{k;\omega_h}$ under the Hamiltonian of the TDHO $H(t)$.
We note that the phase relevant in the on-demand beamsplitter does not coincide with the AA phase because the IF is defined with respect to the memory frequency as in equation \eqref{eq:final_result_interaction_frame}, i.e., the phase $-\omega_h(k+1/2) T$ should be subtracted from the total phase in equation \eqref{eq:PS_equals_iFCandFC}.

\subsection{The Hong--Ou--Mandel effect} \label{subsec:HOM}

\begin{figure}[htbp]
\begin{center}
\includegraphics[width=8.6cm]{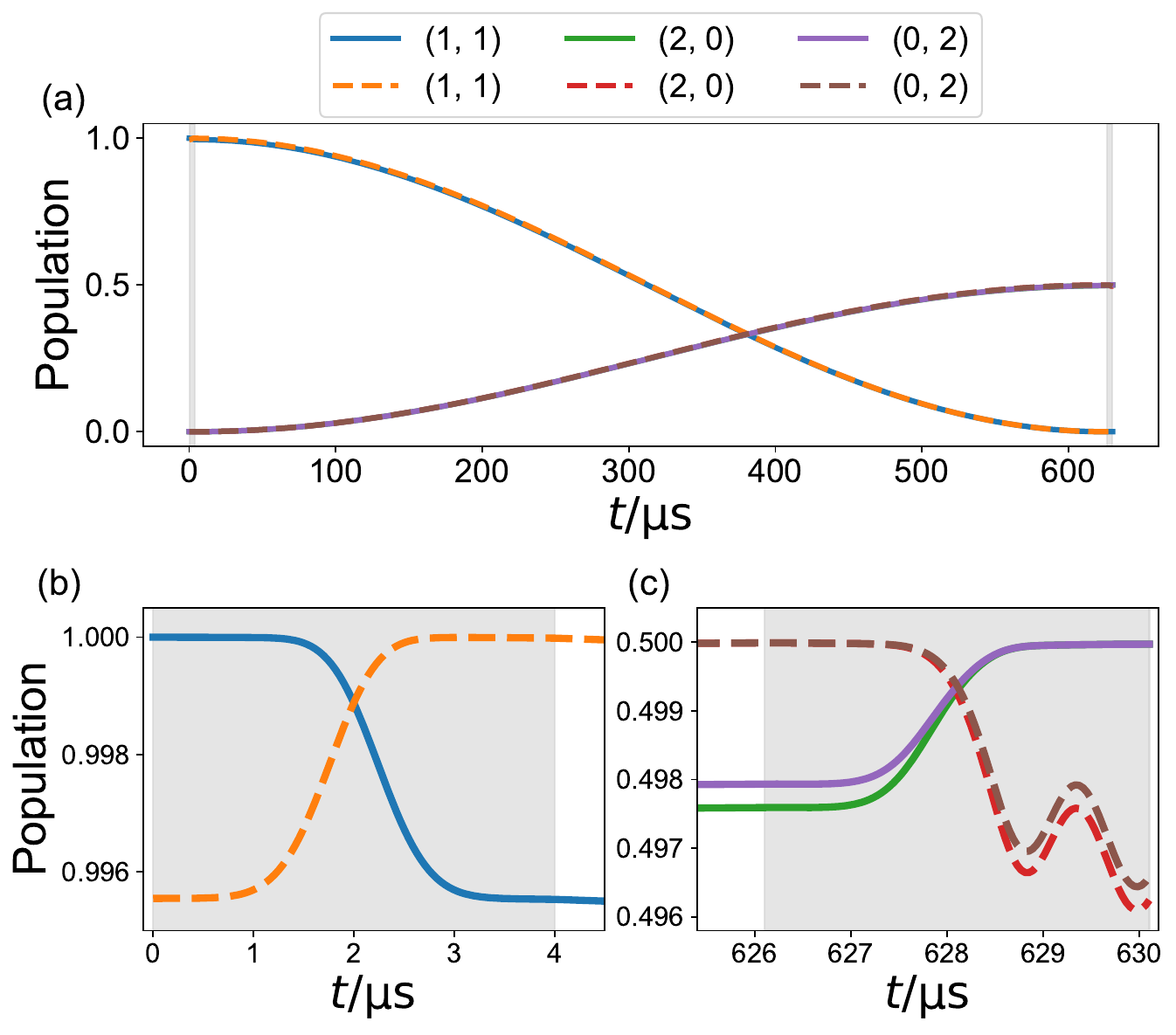}
\caption{(a) The populations of three states relevant in the HOM effect, $\ket{1}_1\ket{1}_0$, $\ket{2}_1\ket{0}_0$, and $\ket{0}_1\ket{2}_0$ in the memory-frequency basis (solid lines) and the gate-frequency basis (dashed lines)
(b) and (c) shows the expanded view during $\FC$ and $i\FC$, respectively.}
\label{fig:HOM}
\end{center}
\end{figure}

The 50:50 beamsplitter is realized by setting $\theta=\pi/4$, which corresponds to $T_{\B}=622\,\mus$. When the input to the 50:50 beamsplitter is $\ket{1}_1\ket{1}_0$, the output is given by
\begin{align}
    \label{eq:HOM}
    U_{\B}(\pi/4)\ket{1}_1\ket{1}_0&=\frac{1}{2}\left(a_1^{\dagger}-\ii a_0^{\dagger}\right)\left(-\ii a_1^{\dagger}+a_0^{\dagger}\right)\ket{0}_1\ket{0}_0 \notag\\
    &=\frac{-\ii}{\sqrt{2}}\left(\ket{2}_1\ket{0}_0 + \ket{0}_1\ket{2}_0 \right),
\end{align}
which is known as the Hong--Ou--Mandel (HOM) effect \cite{HongOuMandel1987}
The HOM effect was demonstrated for trapped ions experimentally \cite{Toyoda2015} using a ``non-on-demand beamsplitter,'' i.e., the frequencies of the two modes were kept the same as $\omega_0=\omega_1=\omega$ throughout the experiment, the phonon hopping started immediately after the initial state $\ket{1;\omega}_1\ket{1;\omega}_0$ was prepared, and the population of each mode was measured after a certain time corresponding to the 50:50 beamsplitter.

When we adopt the on-demand beamsplitter, the output in the IF is obtained using equation \eqref{eq:U_ODB} as
\begin{align}
    \label{eq:HOM_by_ODB}
    &\ket{\psi_{\mathrm{HOM}}}=U_{\ODB}\left(\frac{\pi}{4}\right)\ket{1;\omegal}_1\ket{1;\omegah}_0 \notag \\
    &=U_{i\FC}(-\Theta_{\highmid}+\omegam T_{\B}-\omegah T_3, 
    -\Theta_{\lm}+\omegam T_{\B}-\omegal T_3)\notag\\
    &\hspace{3mm}\times U_{\B}(\pi/4)U_{\FC}(-\Theta_{\highmid}, -\Theta_{\lm})\ket{1;\omegal}_1\ket{1;\omegah}_0 \notag \\
    &\hspace{0mm}=\frac{-\ii}{\sqrt{2}}\left(e^{\ii\phi_{2,0}}\ket{2;\omegal}_1\ket{0;\omegah}_0 + e^{\ii\phi_{0,2}}\ket{0;\omegal}_1\ket{2;\omegah}_0 \right),
\end{align}
where
\begin{align}
    \label{eq:phase_HOM}
    \phi_{2,0}&=\left(\frac{1}{2}\Theta_{\highmid}+\frac{5}{2}\Theta_{\lm}\right)-3\omegam T_{\B}\notag\\
    &\hspace{5mm}+\left(\frac{1}{2}\omegah +\frac{5}{2}\omegal\right)T_3 + \frac{3}{2}\left(\Theta_{\highmid}+\Theta_{\lm}\right), \\
    \phi_{0,2}
    &=\left(\frac{5}{2}\Theta_{\highmid}+\frac{1}{2}\Theta_{\lm}\right)-3\omegam T_{\B}\notag\\
    &\hspace{5mm}+\left(\frac{5}{2}\omegah +\frac{1}{2}\omegal\right)T_3 + \frac{3}{2}\left(\Theta_{\highmid}+\Theta_{\lm}\right).
\end{align}

In the simulation of the HOM effect, we include the hopping term in equation \eqref{eq:Hamiltonian} even during $\FC$ and therefore the unitary operator of $\FC$ can deviate from equation \eqref{eq:FC_operator_as_pureFC_and_phase_shift}.
In figure \ref{fig:HOM}, we show the population of three states involved in the HOM effect, i.e., $\ket{1}_1\ket{1}_0$, $\ket{2}_1\ket{0}_0$, and $\ket{0}_1\ket{2}_0$ in the memory-frequency basis $\{\ket{n;\omegal}_1\ket{n';\omegah}_0\}$ (solid lines) and in the gate-frequency basis $\{\ket{n;\omegam}_1\ket{n';\omegam}_0\}$ (dashed lines).
In figure \ref{fig:HOM} (a), the gray areas at the beginning and the end represent the operation of $\FC$ and $i\FC$, respectively.
The expanded view during $\FC$ shown in figure \ref{fig:HOM} (b) clearly demonstrates the population transfer from the memory-frequency basis $\ket{1;\omegal}_1\ket{1;\omegah}_0$ to the gate-frequency basis $\ket{1;\omegam}_1\ket{1;\omegam}_0$.
The population of $\ket{1;\omegal}_1\ket{1;\omegah}_0$ at the end of $\FC$ as well as that of $\ket{1;\omegam}_1\ket{1;\omegam}_0$ at the beginning of $\FC$ equals to the overlap between them $|\braket{1;\omegal|1;\omegam}\braket{1;\omegah|1;\omegam}|^2=0.994$.
As shown in figure \ref{fig:HOM} (c), the HOM effect can be observed in the gate-frequency basis at the end of $U_{\B}$, i.e., the population of $\ket{2;\omegam}_1\ket{0;\omegam}_0$ and $\ket{0;\omegam}_1\ket{2;\omegam}_0$ are 0.5 at $t=626\,\mus$.
The population of these states are then transferred to the memory-frequency basis, $\ket{2;\omegal}_1\ket{0;\omegah}_0$ and $\ket{0;\omegal}_1\ket{2;\omegah}_0$, at the end of $i\FC$ at $t=630\,\mus$.

By denoting the final state of the simulation as $\ket{\psi_{\mathrm{sim}}}$, the infidelity defined as $1- |\braket{\psi_{\mathrm{HOM}}|\psi_{\mathrm{sim}}}|^2$ is $8.98\times10^{-5}$.
Even if we neglect the hopping term during $\FC$ and $i\FC$, the infidelity is $1.79\times10^{-5}$ due to the numerical error in the TDVP-MPS simulation
and therefore we estimate the effect of the hopping term during $\FC$ and $i\FC$ by the increase of the infidelity, $7.19\times10^{-5}$.

\subsection{SWAP the GKP states} \label{subsec:SWAP_GKP}
\begin{figure*}[htbp]
\begin{center}
\includegraphics[width=17.2cm]{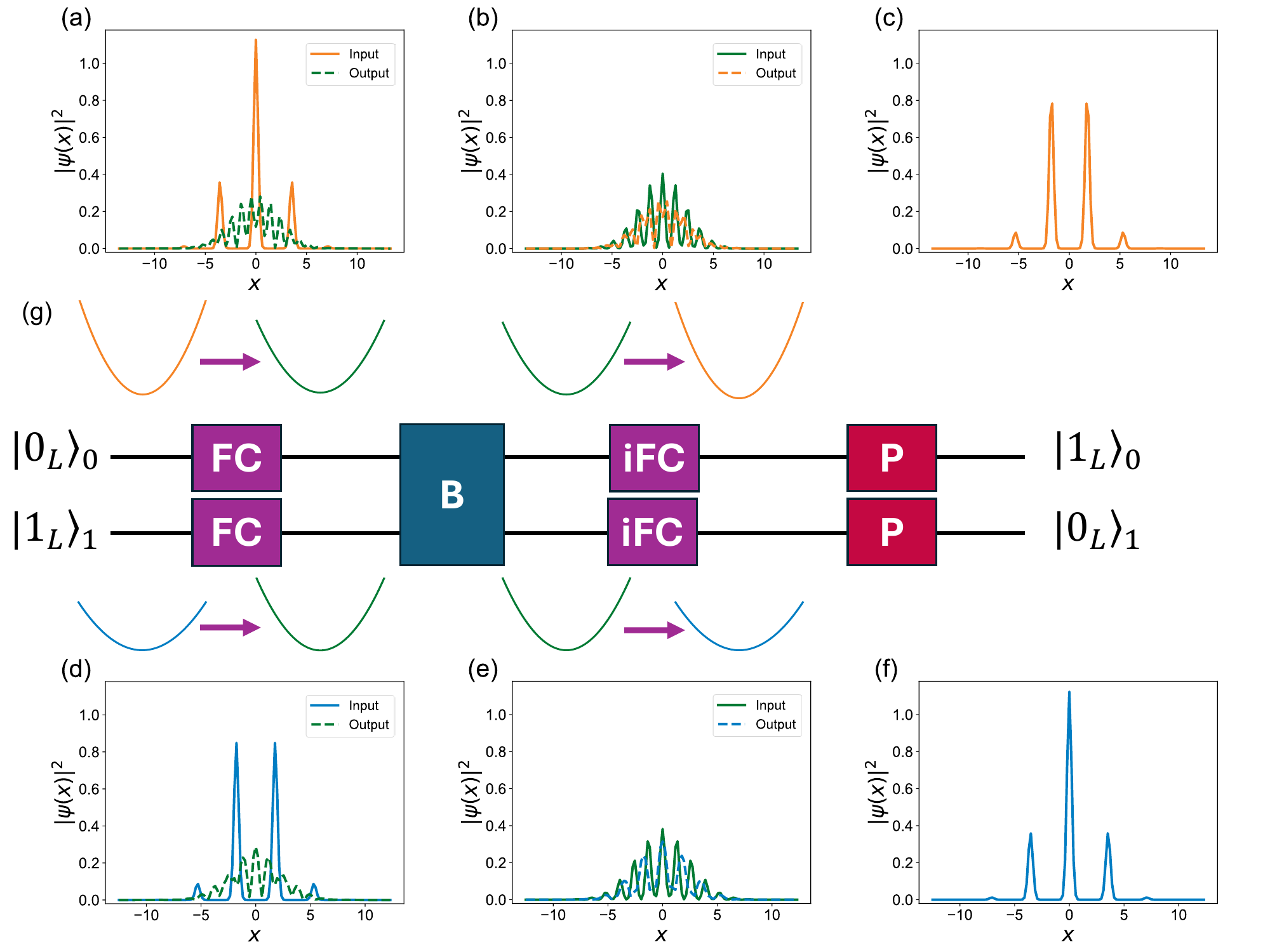}
\caption{The simulation of the $\SWAP$ gate between two GKP states using the on-demand beamsplitter.
The marginal distributions in the position space of each mode are shown in (a)-(f).
The schematic of the on-demand beamsplitter is shown in (g), where the gates are applied from the left to right. 
Colors of the harmonic well and the marginal distributions correspond to the frequencies $\omegah$ (orange), $\omegam$ (green), and $\omegal$ (blue).
The initial state is the product of GKP states $\ket{1_L}_1\ket{0_L}_0$ characterized by the parameters $\Delta=\epsilon=0.3.$
In (a) and (d) ((b) and (e)), the solid lines represent the input to $\FC$ ($i\FC$) while the dashed lines represent the output from $\FC$ ($i\FC$).
(c) and (f) are the output of the phase-shift gate $\PS$, which completes the $\SWAP$ gate.
}
\label{fig:SWAP_GKP}
\end{center}
\end{figure*}

The logical two-qubit gates for the finite-energy GKP codes suffer from the error caused by the finite squeezing of the GKP states.
Rojkov $\etal$ proposed to combine $\SWAP$ gates and qutrit-controlled unitary gates to implement a controlled-Z gate for the finite-energy GKP codes with high fidelity \cite{Rojkov2024}.
Therefore, we demonstrate the $\SWAP$ gate using the on-demand beamsplitter in the following.

First, we double the duration of hopping $T_{\B}$ compared to the 50:50 beamsplitter, i.e., we set $\theta=\pi/2$ and $T_{\B}=1244\,\mus$.
From equation \eqref{eq:B_Schrodinger_pic}, this is not equivalent to the $\SWAP$ gate but is composed of the $\SWAP$ gate and the phase-shift gate as
\begin{align}
    \label{eq:SWAP_mn}
    U_{\B}(\pi/2)\ket{j}_1\ket{k}_0    
    &=\left(-\ii a_0^{\dagger}\right)^j\left(-\ii a_1^{\dagger}\right)^k \ket{0}_1 \ket{0}_0 \notag\\
    &=(-\ii)^{j+k}\ket{k}_1\ket{j}_0\notag\\
    &=\ii U_{\PS}(\pi/2, \pi/2)U_{\SWAP}\ket{j}_1\ket{k}_0.
\end{align}
On the other hand, if we input an arbitrary product state $\ket{\psi;\omegal}_1\ket{\xi;\omegah}_0=\sum_{j}\alpha_j\ket{j;\omegal}_1\sum_k\beta_k\ket{k;\omegah}_0$ to the on-demand beamsplitter with $\theta=\pi/2$, the output is obtained using equation \eqref{eq:U_ODB} as
\begin{align}
    \label{eq:SWAP_by_ODB}
    &U_{\ODB}\left(\frac{\pi}{2}\right)\ket{\psi;\omegal}_1\ket{\xi;\omegah}_0 \notag\\
    &=U_{i\FC}(-\Theta_{\highmid}+\omegam T_{\B}-\omegah T_3, \notag\\
    &\hspace{12mm}-\Theta_{\lm}+\omegam T_{\B}-\omegal T_3) U_{\B}(\pi/2)\notag\\
    &\hspace{5mm}\times\sum_{j,k}\alpha_j e^{\ii\Theta_{\lm}(j+1/2)}\beta_k e^{\ii\Theta_{\highmid}(k+1/2)}\ket{j;\omegam}_1\ket{k;\omegam}_0 \notag \\
    &=\ii U_{i\FC}(-\Theta_{\highmid}+\omegam T_{\B}-\omegah T_3 + \pi/2, \notag\\
    &\hspace{12mm}-\Theta_{\lm}+\omegam T_{\B}-\omegal T_3 + \pi/2) \notag\\
    &\hspace{5mm}\times\sum_{j,k}\alpha_j e^{\ii\Theta_{\lm}(j+1/2)}\beta_k e^{\ii\Theta_{\highmid}(k+1/2)}\ket{k;\omegam}_1\ket{j;\omegam}_0 \notag\\
    &=\ii U_{\PS}(\phi_0, \phi_1) \ket{\xi;\omegal}_1\ket{\psi;\omegah}_0,
\end{align}
where 
\begin{align}
    \label{eq:phases_after_SWAP}
    \phi_0 &=-(\Theta_{\highmid}+\Theta_{\lm})+\omegam T_{\B}-\omegah T_3 + \pi/2 \notag\\
    \phi_1 &=-(\Theta_{\highmid}+\Theta_{\lm})+\omegam T_{\B}-\omegal T_3 + \pi/2.
\end{align}
The phase shift which appears in the last line of equation \eqref{eq:SWAP_by_ODB} can be compensated by applying the phase-shift gate only at the end while in the general case we need to apply the phase-shift gates at both the beginning and the end as shown in equation \eqref{eq:Smatrix_phase_compensated}.
Consequently, the $\SWAP$ gate can be realized as
\begin{align}
    \label{eq:SWAP_phase_compensated}
    U_{\SWAP}&=U_{\PS}(-\phi_0, -\phi_1) U_{\ODB}\left(\frac{\pi}{2}\right)
\end{align}
up to the irrelevant global phase.

We numerically simulate the $\SWAP$ gate acting on the product of finite-energy GKP states, $\ket{\psi_0}=\ket{1_L}_1\ket{0_L}_0\xrightarrow{\SWAP}\ket{\psi_{\SWAP}}=\ket{0_L}_1\ket{1_L}_0$.
The finite-energy GKP states in position space are given by
\begin{align}
    \label{eq:GKP_definition}
    \ket{j_L}=N_{\Delta,\epsilon}\int_{-\infty}^{\infty}dx\sum_{s\in\mathbb{Z}}&e^{-\frac{1}{2}\epsilon^2((2s+j)\sqrt{\pi})^2} \notag\\
    &\times e^{-\frac{1}{2\Delta^2}(x-(2s+j)\sqrt{\pi})^2}\ket{x},
\end{align}
where $N_{\Delta,\epsilon}$ is the normalization constant, $\Delta$ is the width of each peak, and $\epsilon^{-1}$ is the width of the envelope.
We set $\Delta=\epsilon=0.3$, with which the logical infidelity of the controlled-Z gate becomes smaller than $10^{-4}$ \cite{Rojkov2024}.
The sum over $s$ is truncated at $|s|=s_{\max}=6$ and then $N_{\Delta,\epsilon}$ is numerically calculated.
Figure \ref{fig:SWAP_GKP} shows the marginal distribution in the position space of each mode in the memory-frequency basis, $x_{\omegal}$ and $x_{\omegah}$.
Although the populations in the Fock basis are swapped between two modes after $i\FC$, the marginal distribution deviates from that of the logical states because of the phase shift as shown by the dashed lines in figure \ref{fig:SWAP_GKP} (b) and (e).
By applying the appropriate phase-shift gates to each mode, we can obtain the desired logical states as shown in figure \ref{fig:SWAP_GKP} (c) and (f).
The infidelity $1-|\braket{\psi_{\SWAP}|\psi_{\mathrm{sim}}}|^2$ is $2.17\times10^{-3}$.
On the other hand, when we neglect the hopping term during $\FC$ and $i\FC$, the infidelity is $1.21\times10^{-3}$ due to numerical error, 
and therefore we estimate the effect of the hopping term by their difference, $0.96\times10^{-3}$.

\subsection{Effect of undesired hopping during frequency changing at the jump in the sawtooth configuration} \label{subsec:SWAP_GKP_4modes}
The on-demand beamsplitter between the modes at the large jump in the sawtooth configuration needs careful analysis.
We consider the 3rd and the 4th modes in figure \ref{fig:sawtooth} and so the frequencies involved in $\FC$ are given as $\omegah=\omega_4=\omega'_0$, $\omegal=\omega_3=\omega'_3$, and $\omegam=\omega_{1,2}$.
In order to keep the rate of $\FC$ as $2\pi\times50\,\kHz/\mus$, the duration of $\FC$ is set to $12\,\mus$.
The width of the error function is set to $\sigma=4.5$ to reduce undesired hopping during $\FC$.

Because the simulation of all modes involved in the sawtooth configuration is infeasible, we use subsets of the modes and estimate the error induced by undesired hopping.
In addition, the computational cost is further reduced by lowering the energy of GKP states, i.e., we set $\Delta=0.6$ and $\epsilon=0.7$.
This allows us to reduce the cost of representing the state in the 1st quantization; the number of grid points is 64 and the maximum of the bond dimension is 20.

First, we consider the subset consisting of the 3rd and the 4th modes and simulate the $\SWAP$ gate for the initial state $\ket{1_L}_4\ket{0_L}_3$.
The infidelity is $9.33\times10^{-4}$, which originates from the numerical error and the undesired hopping during $\FC$ and $i\FC$ as we have seen in section \ref{subsec:SWAP_GKP}.

When we add the 2nd and the 5th modes to the subset and we set the initial state as $\ket{0_L}_5\ket{1_L}_4\ket{0_L}_3\ket{0_L}_2$,
the infidelity of the $\SWAP$ gate between the 3rd and the 4th modes is $1.04\times10^{-3}$.
The increase of the infidelity is due to the cross-talk between NN modes, i.e., hopping between the 2nd and the 3rd modes as well as the 4th and the 5th modes during $\FC$ and $i\FC$ as indicated by curly arrows in figure \ref{fig:sawtooth}.
If we can include other modes in the simulation, the cross-talks between non-NN modes should be taken into account, e.g., between the 3rd and the 6th modes as well as the 1st and the 4th modes.
However, the effect of such non-NN cross-talks should be much smaller because the hopping rate decreases as $\propto d_{j,k}^{-3}$.

\begin{figure}[htbp]
\begin{center}
\includegraphics[width=8.6cm]{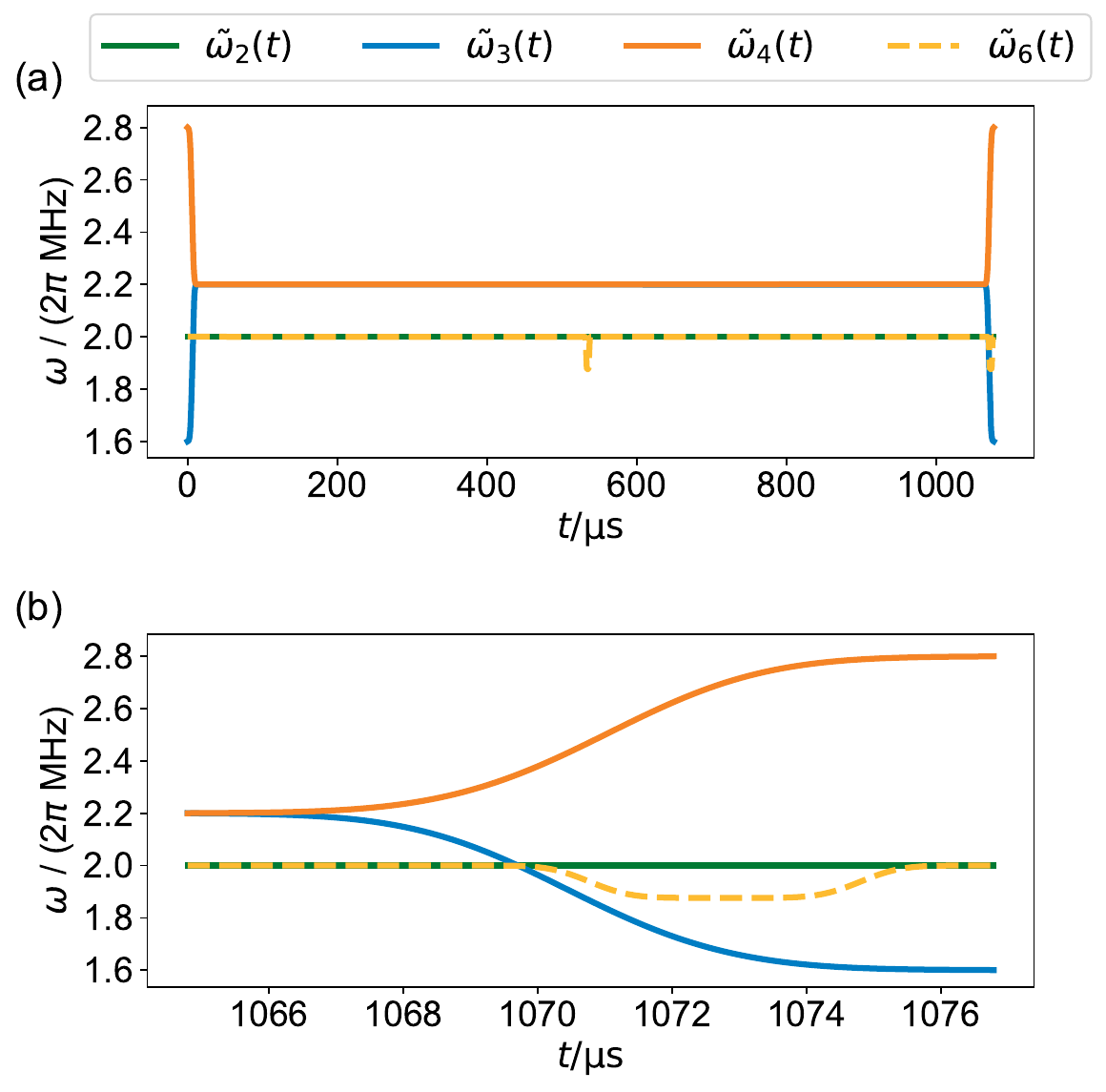}
\caption{Time dependent frequencies of four modes involved in SWAP and C3PO.
(a) Resonant hopping during $\tilde{\omega}_3=\tilde{\omega}_4$ implements the on-demand beamsplitter
while resonant hopping during $\tilde{\omega}_2=\tilde{\omega}_6$ causes error. 
The crossing of $\tilde{\omega}_2$ and $\tilde{\omega}_3$ causes the NN cross-talk and the crossing of $\tilde{\omega}_3$ and $\tilde{\omega}_6$ causes the non-NN cross-talk.
The tiny dips of $\tilde{\omega}_6$ at the middle and at the end correspond to the application of $\PS_{-\pi}$.
(b) The expanded view of (a) around the end of the simulation.
}
\label{fig:omega_SWAP_C3PO}
\end{center}
\end{figure}

Then, we consider a subset of the 2nd, the 3rd, the 4th, and the 6th modes.
With this subset, we can investigate the effect of resonant hopping between the 2nd and the 6th modes in addition to the cross-talk between the 2nd and the 3rd as well as the 3rd and the 6th modes.
For the initial state $\ket{0_L}_6\ket{1_L}_4\ket{0_L}_3\ket{0_L}_2$, the infidelity of the $\SWAP$ gate between the 3rd and the 4th modes is $3.36\times10^{-3}$.
The resonant hopping between the 2nd and the 6th modes can be mitigated by applying C3PO \cite{Nishi2025C3PO}; we apply $(-\pi)$-phase shift gates $\PS_{-\pi}$ to the 6th mode at the middle and the end of the on-demand beamsplitter and obtain the smaller infidelity $2.32\times10^{-3}$.
The frequency of each mode is shown in figure \ref{fig:omega_SWAP_C3PO}.
From equation \eqref{eq:PS_equals_iFCandFC}, $\PS_{-\pi}$ is decomposed into $\FC$ and $i\FC$.
As shown in figure \ref{fig:omega_SWAP_C3PO}, $\PS_{-\pi}$ is implemented by downward $\FC$ and its inverse so that the frequency of the 6th mode is kept away from the gate frequency $\omega_{1,2}$ used in the on-demand beamsplitter.
We choose the parameters specifying $\FC$ in $\PS_{-\pi}$ as $T_{\FC}=4\,\mus,\,\sigma=6,\omega_i=\omega'_2,\omega_f=\omega'_2-2\pi\times124\,\kHz$ and similarly for $i\FC$.
The crossing of $\tilde{\omega}_2$ and $\tilde{\omega}_3$ as well as that of $\tilde{\omega}_3$ and $\tilde{\omega}_6$ corresponds to the cross-talk error.

\section{Conclusion}
We have proposed a method for implementing an on-demand beamsplitter for trapped-ion quantum computers by combining frequency changing, phonon hopping, and phase-shift gates.
Based on the analytical solution of the time-dependent harmonic oscillator, we have derived the transformation matrix of the on-demand beamsplitter in the Heisenberg picture.
Because the transformation matrix is obtained by neglecting the hopping terms during the operation of the frequency changing, the time-evolution under the realistic Hamiltonian can deviate from the analytical results.
In order to evaluate the effect of the hopping terms, we have numerically solved the TDSE using TDVP-MPS method.
The HOM effect and the SWAP of the GKP states have been successfully demonstrated.
Due to numerical error, the simulated results deviate from the analytical solution even when we neglect the hopping terms in the simulation.
Although the inclusion of the hopping terms increases the infidelity, it is difficult to separate the effects of the numerical error and the physical errors from the hopping terms.
For more detailed analysis of the effect of hopping, the TDSE solver should be improved.
The frequency change and the phase-shift gate could be implemented experimentally by modulating the DC voltage on the electrodes of a surface electrode trap \cite{Mielenz2016}.
The parameters used in the numerical simulation are comparable to those used in the experiment \cite{Mielenz2016} and thus our proposal should be experimentally feasible.
Moreover, because the analytical formula does not rely on the adiabatic approximation, it is possible to use a faster frequency change than our simulation.
For large number $M$ of modes, we have proposed the sawtooth configuration; when we can use $N$ memory frequency, any two resonant modes are separated by $Nd$.
We have also shown that the slow hopping between those resonant modes can be mitigated by applying the dynamical decoupling.
Alternatively, we can mitigate the resonant hopping by using the shifted sawtooth configuration.
In both cases, we could adopt an optical tweezer to modify the trap potential \cite{Mazzanti2021} because the amount of frequency change needed can be small.
We expect that the proposed method will contribute to the scaling up of the CV-encoded trapped-ion quantum computers as well as the analog quantum simulator.

\begin{acknowledgments}
The author thanks Kaoru Yamanouchi for his valuable comments on the present study, Takashi Mukaiyama and Tomoya Okino for their helpful discussion on experimental aspects, Yutaka Shikano for the comments on the geometric phase quantum computation, and Kenji Toyoda and Ting Rei Tan for stimulating discussion.
This work was supported by JST-CREST Grant No. JPMJCR23I7.
\end{acknowledgments}

\appendix
\section{Hamiltonian of a chain of trapped ions}\label{app:Hamiltonian}
We consider $\nMode$ trapped ions and their local modes along $X$ axis and assume that, if we neglect the Coulomb coupling, the secular frequency of each mode is given by $\omega_{0,j}$.
Then, the total Hamiltonian for $\nMode$ harmonic oscillators interacting through the Coulomb coupling is given by \cite{Nishi2025C3PO}
\begin{align}
    \label{eqApp:bare_Hamiltonian}
    H_0&=\sum_j\left(\frac{P_j^2}{2m} + \frac{m X_j^2}{2}\left(\omega_{0,j}^2 - \sum_{k\neq j}\frac{e^2}{4\pi\epsilon_0 d_{j,k}^3m} \right)\right) \notag\\
    &\hspace{5mm}+\sum_{j>k} \frac{e^2}{4\pi\epsilon_0 d_{j,k}^3} X_jX_k,
\end{align}
where $m$ is the mass of an ion and $d_{j,k}$ is the distance between the equilibrium position of the $j$th and the $k$th modes.
In order to simplify the notation, we redefine the secular frequency of the $j$th mode as $\omega_j\equiv\sqrt{\omega_{0,j}^2 - \sum_{k\neq j}e^2/(4\pi\epsilon_0 d_{j,k}^3m)}$.
By defining the scaled coordinate and momentum as $x_j=X_j/x_{0,j}$ and $p_j=P_j/p_{0,j}$, respectively, with $x_{0,j}=\sqrt{\hbar/(m\omega_j)}$ and $p_{0,j}=\sqrt{\hbar m\omega_j}$, Eq. \ref{eqApp:bare_Hamiltonian} is rewritten as
\begin{align}
    H_0&=\sum_j\frac{\hbar \omega_j}{2} \left( p_j^2+x_j^2 \right) 
    +\sum_{j>k} \hbar\kappa_{j,k} x_jx_k \notag \\
    &\simeq \sum_j\frac{\hbar \omega_j}{2} \left( p_j^2+x_j^2 \right) 
    +\sum_{j>k}\frac{\hbar\kappa_{j,k}}{2} (x_jx_k + p_jp_k) \label{eqApp:RWA_Hamiltonian_scaled_xp}\\
    &=\sum_j\hbar\omega_j\left(a_j^{\dagger}a_j + \frac{1}{2}\right) 
    +\sum_{j>k}\frac{\hbar\kappa_{j,k}}{2}\left(a_j^{\dagger}a_k+a_ja_k^{\dagger}\right),\label{eqApp:RWA_Hamiltonian_2nd_quantization}
\end{align}
where $\kappa_{j,k}=e^2/(4\pi\epsilon_0 d_{j,k}^3m\sqrt{\omega_j\omega_k})$.
The second line is obtained by applying the rotating-wave approximation and the third line is obtained by defining $x_j=(a^{\dagger}_j + a_j)/\sqrt{2}$ and $p_j=\ii(a^{\dagger}_j - a_j)/\sqrt{2}$.

For the FC, we can modulate the bare frequency as $\tilde{\omega}_{0,j}(t)$, which satisfies $\tilde{\omega}_{0,j}(0)=\omega_{0,j}$ and we define $\tilde{\omega}_j(t)\equiv\sqrt{\tilde{\omega}_{0,j}(t)- \sum_{k\neq j}e^2/(4\pi\epsilon_0 d_{j,k}^3m)}$ as the time-dependent frequency of the TDHO.
Therefore, the Hamiltonian of the TDHO can be written as
\begin{align}
    H(t)&=\sum_j\left(\frac{P_j^2}{2m} + \frac{m}{2}\tilde{\omega}_j^2(t)X_j^2\right) 
   +\sum_{j>k} \frac{e^2}{4\pi\epsilon_0 d_{j,k}^3} X_jX_k \notag\\
   &=H_0 + \sum_j{\frac{\hbar\Omega_j^2(t)}{2\omega_j}x_j^2} \label{eqApp:Hamiltonian_TDHO_1st}\\
   &=H_0 + \sum_j{\frac{\hbar\Omega_j^2(t)}{4\omega_j}(a^{\dagger}_j + a_j)^2}, \label{eqApp:Hamiltonian_TDHO_2nd}
\end{align}
where we defined the difference of the squared frequency as $\Omega_j^2(t)\equiv \tilde{\omega}_j^2(t) - \omega_{j}^2=\tilde{\omega}_{0,j}^2(t) - \omega_{0,j}^2$.

\section{Basis change}\label{app:Basis_change}
When we consider the TDHO with $H(t)=P^2/(2m) + m\tilde{\omega}^2(t) X^2/2$ and the time evolution from $\tilde{\omega}(t_i)=\omega_i$ to $\tilde{\omega}(t_f)=\omega_f$,
the scaled coordinates should be defined at $t_i$ and $t_f$ as $x_{\omega_i}=\sqrt{m\omega_i/\hbar}X$ and $x_{\omega_f}=\sqrt{m\omega_f/\hbar}X$, respectively,
and are related as
\begin{align}
    x_{\omega_f}&=\sqrt{\frac{\omega_f}{\omega_i}}x_{\omega_i}\equiv r_{\omega} x_{\omega_i}, \label{eqApp:x_basis_change}\\
    p_{\omega_f}&=\sqrt{\frac{\omega_i}{\omega_f}}p_{\omega_i}\equiv r_{\omega}^{-1} p_{\omega_i}.
\end{align}
By denoting the lowering operator at each time as $a_{\omega_i}$ and $a_{\omega_f}$,
their relation can be obtained as
\begin{align}
    \label{eqApp:a_basis_change}
    a_{\omega_i}&=\frac{1}{\sqrt{2}}\left(x_{\omega_i}+\ii p_{\omega_i}\right)\notag\\
    &=\frac{1}{2}\left(r_{\omega}^{-1}\left(a^{\dagger}_{\omega_f}+a_{\omega_f}\right)
    - r_{\omega}\left(a^{\dagger}_{\omega_f}-a_{\omega_f}\right)\right) \notag\\
    &=\frac{r_{\omega}+r_{\omega}^{-1}}{2}a_{\omega_f} - \frac{r_{\omega}-r_{\omega}^{-1}}{2}a^{\dagger}_{\omega_f} \notag\\
    &=a_{\omega_f}\cosh{r} - a^{\dagger}_{\omega_f}\sinh{r},
\end{align}
where we defined $r=\log{r_{\omega}}$.
Therefore, the basis change can be regarded as the squeezing with $r$ the squeezing strength.

\section{Lewis--Riesenfeld theory}\label{app:LR_theory}
The analytical solution of the TDHO was derived in the Heisenberg picture by Lewis and Riesenfeld \cite{LewisRiesenfeld1969} as we summarize below.
Note that we adopt the notation used in Ref. \cite{Lau&James2012}.
If $\ket{\psi}$ is the solution of the TDSE $\ii\hbar\partial_t\ket{\psi}=H(t)\ket{\psi}$,
$I(t)\ket{\psi}$ is also the solution, where $I(t)$ is a Hermitian operator satisfying 
\begin{align}
    \label{eqApp:Invariant_general}
    \frac{\partial I(t)}{\partial t} + \frac{1}{\ii\hbar}[I(t),\,H(t)]=0,
\end{align}
and we denote $I(t)$'s eigenvalues and eigenstates as $\lambda_n$ and $\ket{\lambda_n,t}$, respectively.
For the TDHO specified in Appendix \ref{app:Basis_change}, by defining the invariant as 
\begin{align}
    \label{eqApp:Invariant_TDHO}
    I(t) = \frac{\left(bp-m\dot{b}x\right)^2}{2m} + \frac{m}{2}\left(\frac{\omega_i}{b}\right)^2x^2,
\end{align}
where $b(t)$ satisfies
\begin{align}
    \ddot{b}+\tilde{\omega}^2(t)b-\frac{\omega_i^2}{b^3}&=0 \label{eqApp:b} \\
    \Longleftrightarrow\tilde{\omega}(t)&=\sqrt{\left(\frac{\omega_i^2}{b^3}-\ddot{b}\right)/b}, \label{eqApp:omega(t)_by_b(t)}
\end{align}
the lowering and the raising operators for the eigenstates of $I(t)$ can be defined respectively as
\begin{align}
    A(t)\ket{\lambda_n,t} = \sqrt{n}\ket{\lambda_{n-1},t}, \label{eqApp:TD_lowering}\\
    A^{\dagger}(t)\ket{\lambda_n,t} = \sqrt{n+1}\ket{\lambda_{n+1},t}.\label{eqApp:TD_rasing}
\end{align}
The eigenstates of $I(t)$ are related to the solution of the TDSE as $\ket{\lambda_n,t}e^{\ii(n+1/2)\Theta(t)}$, where
\begin{align}
    \label{eqApp:Theta}
    \Theta(t) = -\int_{t_i}^{t}{\frac{\omega_i}{b^2(t')}dt'}.
\end{align}
We choose the initial condition of $\eqref{eqApp:b}$ as $b(t_i)=1$ and $\dot{b}(t_i)=0$ so that $I(t_i)=H(t_i)$ and $A(t_i)=a_{\omega_i}$ are satisfied.
At the end of the propagation, the lowering operator is given as
\begin{align}
    \label{eqApp:A_tf}
    A(t_f) = \eta(t_f)a_{\omega_f} + \zeta(t_f)a^{\dagger}_{\omega_f},
\end{align}
where
\begin{align}
    \eta(t)&=\frac{1}{2}\sqrt{\frac{\omega_i}{\omega_f}}\left(\frac{1}{b}+\frac{\omega_f}{\omega_i}b-\ii\frac{\dot{b}}{\omega_i}\right), \label{eqApp:eta}\\
    \zeta(t)&=\frac{1}{2}\sqrt{\frac{\omega_i}{\omega_f}}\left(\frac{1}{b}-\frac{\omega_f}{\omega_i}b-\ii\frac{\dot{b}}{\omega_i}\right).\label{eqApp:zeta}
\end{align}

\section{Frequency changing}\label{app:FC}
In order to derive the time evolution of the lowering operator in the Heisenberg picture, we expand the time-evolution operator as
\begin{align}
    \label{eqApp:U(t)}
    U(t_f,t_i) = \sum_{n=0}^{\infty}{e^{\ii(n+1/2)\Theta(t_f)}\ket{\lambda_n,t_f}\bra{\lambda_n,t_i}}.
\end{align}
By using Eq. \eqref{eqApp:TD_lowering}, $A(t_f)$ can be related to $a_{\omega_i}$ as
\begin{align}
    \label{eqApp:A(tf)_ai}
    &U(t_f,t_i)A(t_i)U^{\dagger}(t_f,t_i) \notag\\
    &=U(t_f,t_i)\sum_{n=0}^{\infty}{e^{-\ii(n+1/2)\Theta(t_f)}\sqrt{n}\ket{\lambda_{n-1},t_i}\bra{\lambda_n,t_f}} \notag\\
    &=e^{-\ii\Theta(t_f)}\sum_{n=0}^{\infty}{\sqrt{n}\ket{\lambda_{n-1},t_f}\bra{\lambda_n,t_f}}
    =e^{-\ii\Theta(t_f)}A(t_f)\notag\\
    &\Longrightarrow A(t_f)=e^{\ii\Theta(t_f)}U(t_f,t_i)a_{\omega_i}U^{\dagger}(t_f,t_i).
\end{align}
Combining this with Eq. \eqref{eqApp:A_tf}, we obtain
\begin{align}
    \label{eqApp:A_eta_zeta_UaU}
    A(t_f)&=\eta(t_f) a_{\omega_f} + \zeta(t_f) a^{\dagger}_{\omega_f}\notag\\
    &=e^{\ii\Theta(t_f)}U(t_f,t_i)a_{\omega_i}U^{\dagger}(t_f,t_i),
\end{align}
from which we can relate the time evolution of $a_{\omega_f}$ with $a_{\omega_i}$ as
\begin{align}
    &\eta(t_f)a_{\omega_f}(t_f) + \zeta(t_f)a^{\dagger}_{\omega_f}(t_f) = e^{\ii\Theta(t_f)}a_{\omega_i} \notag\\
    &\Longrightarrow a_{\omega_f}(t_f)=\eta^* e^{\ii\Theta(t_f)}a_{\omega_i} - \zeta e^{-\ii\Theta(t_f)}a^{\dagger}_{\omega_i}, \label{eqApp:af_by_ai}
\end{align}
where we used $a_{\omega_f}(t)=U^{\dagger}(t,t_i)a_{\omega_f}U(t,t_i)$.
By denoting the lowering and the raising operators by a vector $\bm{a}_{\omega}=\begin{bmatrix}
    a_{\omega}, a^{\dagger}_{\omega}
\end{bmatrix}^{\top}$, Eq. \eqref{eqApp:af_by_ai} can be summarized using a transformation matrix $S_{if}$ as $\bm{a}_{\omega_f}(t_f)=S_{if}\bm{a}_{\omega_i}$, where 
\begin{align}
    \label{eqApp:Smatrix_af_by_ai}
    S_{if} =
    \begin{bmatrix}
        \eta^*e^{\ii\Theta(t_f)} & -\zeta e^{-\ii\Theta(t_f)}\\   
        -\zeta^* e^{\ii\Theta(t_f)} & \eta e^{-\ii\Theta(t_f)}
    \end{bmatrix}.
\end{align}
Finally, by applying Eq. \eqref{eqApp:a_basis_change} to Eq. \eqref{eqApp:af_by_ai}, the time evolution of $a_{\omega_i}$ can be obtained as
\begin{align}
    \label{eqApp:ai(tf)}
    a_{\omega_i}(t_f)&=a_{\omega_f}(t_f)\cosh{r} - a^{\dagger}_{\omega_f}(t_f)\sinh{r} \notag\\
    &=\left(\eta^* \cosh{r}+\zeta^* \sinh{r}\right) e^{\ii\Theta(t_f)}a_{\omega_i} \notag\\
    &\hspace{5mm}- \left(\eta \sinh{r}+\zeta \cosh{r}\right) e^{-\ii\Theta(t_f)}a^{\dagger}_{\omega_i}.
\end{align}

\section{Adiabatic condition}\label{app:adiabatic_condition}
Although the adiabatic condition does not need to be satisfied for $\FC$, the parameters adopted in \ref{sec:Results} are in the adiabatic regime as shown below.
For a system driven by a non-periodic time-dependent potential, the adiabatic condition is given as \cite{Comparat2009}
\begin{align}
    \label{eqApp:adiabatic_condition_general}
    \sum_{m\neq n}\frac{\hbar|\bra{m;t}\dot{H}\ket{n;t}|}{|E_m(t)-E_n(t)|^2} \ll 1,
\end{align}
where $E_n(t)$ and $\ket{n;t}$ are the $n$th eigen energy and the eigen vector of the instantaneous Hamiltonian, i.e., $H(t)\ket{n;t}=E_n(t)\ket{n;t}$.
When we consider a single mode, the Hamiltonian of $\FC(\omega_i\xrightarrow{}\omega')$ and its time derivative are given by
\begin{align}
    H(t)&=\frac{\hbar \omega_i}{2} \left( p^2+x^2 \right) 
    + \frac{\hbar(\tilde{\omega}^2(t)-\omega^2_i)}{2\omega_i}x^2 \label{eqApp:H_single_mode_FC} \\
    \dot{H}&=\frac{\hbar\tilde{\omega}(t)\dot{\tilde{\omega}}(t)}{\omega_i}x^2,
\end{align}
and the instantaneous eigenenergy is given by $E_n(t)=\hbar\tilde{\omega}(t)(n+1/2)$.
In order to calculate $|\bra{m;t}\dot{H}\ket{n;t}|\propto |\bra{m;t}x^2\ket{n;t}|$, we need to change the basis using Eq. \eqref{eqApp:x_basis_change} as
\begin{align}
    \label{eqApp:x_to_xt}
    x = \sqrt{\frac{\omega_i}{\tilde{\omega}(t)}}x_t.
\end{align}
By defining the lowering and the raising operators with respect to $\ket{n;t}$, Eq. \eqref{eqApp:adiabatic_condition_general} can be rewritten as
\begin{align}
    \label{eqApp:adiabatic_condition_TDHO}
    \sum_{m\neq n}\frac{|\dot{\omega}\bra{m;t}(a_t^{\dagger} + a_t)^2\ket{n;t}|}{2\tilde{\omega}^2(t)(m-n)^2}\ll 1.
\end{align}
In the numerator of Eq. \eqref{eqApp:adiabatic_condition_TDHO}, only the terms satisfying $|m-n|=2$ are finite and therefore we only require the terms in the sum in Eq. \eqref{eqApp:adiabatic_condition_TDHO} with large $n$ to satisfy the following,
\begin{align}
    \label{eqApp:adiabatic_condition_simplified}
    \frac{\sqrt{(n+1)(n+2)}|\dot{\tilde{\omega}}(t)|}{8\tilde{\omega}^2(t)} \simeq 
    \frac{n|\dot{\tilde{\omega}}(t)|}{8\tilde{\omega}^2(t)} \ll 1.
\end{align}

For a smooth function such as $\tilde{\omega}(t)$ shown in Fig. \ref{fig:b_and_omega}, $|\dot{\tilde{\omega}}(t)|$ becomes its maximum at $t=T_{\FC}/2$.
For $b(t)$ defined using equation \eqref{eq:fu_erf} or \eqref{eq:fd_erf}, $\ddot{b}(T_{\FC}/2)=0$ is satisfied and so we approximate Eq. \eqref{eqApp:omega(t)_by_b(t)} as $\tilde{\omega}(t)\simeq\omega_i/b^2(t)$ around $t=T_{\FC}/2$.
Consequently, the time derivative is given as
\begin{align}
    \label{eqApp:omega_dot}
    \dot{\tilde{\omega}}\left(\frac{T_{\FC}}{2}\right)=-2\omega_i \dot{b}b^{-3}=-2\omega_i \left(\mp\frac{g\sigma}{\sqrt{\pi}T_{\FC}}\right)\left(1\mp\frac{g}{2}\right)^{-3},
\end{align}
where the minus (plus) sign in parentheses is for the upward (downward) frequency changing.
For the parameters specified in \ref{sec:Results}, the left-hand side of Eq. \eqref{eqApp:adiabatic_condition_simplified} is obtained as $6\times10^{-4}n$ for $\FC(\omegah\xrightarrow{}\omegam)$ and $7\times10^{-4}n$ for $\FC(\omegal\xrightarrow{}\omegam)$.
Therefore, the adiabatic condition is satisfied even when $n$ is as large as a few hundred.
Further simplification can be made when $g$ is small.
Because $b^{-3}\simeq1$ for small $g$, Eq. \eqref{eqApp:adiabatic_condition_simplified} can be rewritten as
\begin{align}
    \label{eqApp:addiabatic_condition_final}
    \frac{|g|\sigma}{4\sqrt{\pi}\omega(T_{\FC}/2)T_{\FC}}n \ll 1,
\end{align}
where $\tilde{\omega}(T_{\FC}/2)T_{\FC}/(2\pi)\simeq9$ is the number of oscillation cycles within $T_{\FC}$.
Therefore, for a small value of $|g|$, even when we decrease the duration down to the single oscillation period, i.e., $\tilde{\omega}(T_{\FC}/2)T_{\FC}/(2\pi)\simeq1$, the adiabatic condition is still valid if $n$ is several dozen or smaller.



%

\end{document}